\documentclass[%
 reprint,
 amsmath,amssymb,
 aps,
]{revtex4-2}

\usepackage{graphicx}
\usepackage{dcolumn}
\usepackage{bm}
\usepackage[inline]{enumitem}
\usepackage{braket}
\usepackage{siunitx}
\usepackage{booktabs}

\usepackage{hyperref}

\usepackage{amsmath}
\usepackage{bbm}

\usepackage{tabularx}
\usepackage{multirow}
\usepackage{makecell}
\usepackage{subfig} 

\usepackage{algorithm}
\usepackage{algpseudocode}

\usepackage[a4paper,hmargin=2.8cm,vmargin=2.0cm,includeheadfoot]{geometry}

\usepackage{tikz}
\usetikzlibrary{arrows.meta, positioning}
\usepackage{hyperref}

\DeclareMathOperator*{\argmax}{arg\,max}

\begin{document}

\title{Bayesian Optimization for Quantum Error-Correcting Code Discovery}
\author{Yihua Chengyu}

\author{Richard Meister}%
\author{Conor Carty}
\author{Sheng-Ku Lin}
\author{Roberto Bondesan}
\email{r.bondesan@imperial.ac.uk}
\affiliation{%
 Department of Computing, Imperial College London, 180 Queen's Gate, London SW7 2AZ, United Kingdom
}%

\begin{abstract}
Quantum error-correcting codes protect fragile quantum information by encoding it redundantly, but identifying codes that perform well in practice with minimal overhead remains difficult due to the combinatorial search space and the high cost of logical error rate evaluation. 
We propose a Bayesian optimization framework to discover quantum error-correcting codes that improves data efficiency and scalability with respect to previous machine learning approaches to this task.
Our main contribution is a multi-view chain-complex neural embedding that allows us to predict the logical error rate of quantum LDPC codes without performing expensive simulations.
Using bivariate bicycle codes and code capacity noise as a testbed, our algorithm discovers a high-rate code 
$[[144,36]]$ that achieves competitive per-qubit error rate compared to the gross code, as well as a low-error code 
$[[144,16]]$ that outperforms the gross code in terms of error rate per qubit. These results highlight the ability of our pipeline to automatically discover codes balancing rate and noise suppression, while the generality of the framework enables application across diverse code families, decoders, and noise models. 

\end{abstract}

\maketitle

\section{Introduction}

Quantum error correction is critical to implementing fault-tolerant quantum algorithms with provable quantum advantage, such as Shor's factoring algorithm \cite{shor1999polynomial}.

Several recent experiments have demonstrated the realisation of small scale quantum memories and logical computations \cite{google2025quantum,reichardt2024demonstration,brock2025quantum,putterman2025hardware}. 
To go beyond these proof-of-principle results, 
better hardware needs to be developed alongside improved quantum error correction protocols that minimize resource overhead.

Quantum error correction protects logical information in quantum codes. These can be classified in terms of their length, rate, and distance, which correspond to the number of physical qubits, the number of logical qubits, and the maximum number of qubits that may be affected by a correctable error event, respectively.
While the code rate is controlled by choosing the dimension of the code space, computing the distance is an NP-hard problem \cite{Kapshikar_2023}, 
complicating the evaluation and characterization of quantum codes.
The traditional approach to quantum code design relies on human ingenuity and results from classical coding theory.
This has led to remarkable theoretical achievements, such as the relatively recent discovery of good quantum codes with rate and distance that scale linearly with code length \cite{panteleev2022asymptotically,leverrier2022quantumtannercodes}.

Asymptotic guarantees on encoding rate and distance, however, do not immediately translate into practical performance in a setting with limited number of qubits.
Moreover, computing the logical error rate of a given code -- the main figure of merit used to compare codes -- requires Monte Carlo simulations of the hardware, which in turn depend on a noise model and a decoder. Often, $10^{5}$–$10^{7}$ trials for each physical error rate are required to obtain  statistically significant estimates of the logical error rate, which makes the numerical evaluation of the performance of a given code computationally expensive.

To overcome these challenges, recent progress towards designing practically useful quantum codes formulates the task as an optimization problem. The objective function then is a figure of merit of the code, such as its distance or logical error rate, and the search space is a given set of quantum codes.
For example, Ref.~\cite{bravyi2024high} performed a brute force search over bicycle bivariate codes to find good codes such as the gross code.
But brute force is limited to scanning a small subspace of the possible codes, and several recent works tackle the optimization problem using machine learning methods to improve scalability
\cite{nautrup2019optimizing,su2023discovery,mauron2024optimization,olle2024simultaneous,he2025discoveringhighlyefficientlowweight,webster2024engineering,freire2025optimizing,cao2022quantum,guerrero2025game,he2025co,lanka2025optimizingcontinuoustimequantumerror}.
This follows related trends of applying machine learning in quantum control \cite{Gebhart_2023,niu2019universal,PhysRevLett.129.050507},
error decoding \cite{Krastanov_2017,Meinerz_2022,egorov2023end,bausch2024learning} and quantum circuit compilation \cite{fosel2021quantum,ruiz2024quantumcircuitoptimizationalphatensor}.

Despite previous works on 
machine learning for code discovery having demonstrated the potential of reinforcement learning (RL) and related methods to optimize stabilizer codes or assemble new ones from modular building blocks, the applicability and scalability of these methods has been limited so far.
For details about existing literature, see Appendix \ref{app:related_work}.

The challenges encountered in optimization over quantum codes arise  for two main reasons.
First, the search space is discrete and very large -- the set of possible stabilizer codes grows super-exponentially with the number of qubits \cite{Gross_2006}.
Second, evaluating useful objective metrics, e.g. the code distance or the logical error rate, 
is expensive, as explained above.
This latter property is particularly problematic when using model-free RL -- as in all the related works cited above -- because the agent lacks an explicit model of the reward and therefore requires many evaluations to converge to a good solution.

To go beyond the state of the art in machine learning for quantum code discovery, in this work, we employ Bayesian optimization (BO), which creates a cheap surrogate model of the objective, and uses that to explore the search space in a data-efficient way.
For this reason, BO is typically the go-to method for optimizing expensive black-box cost functions, and has been applied successfully in various domains, from hyperparameter optimization of neural networks~\cite{NIPS2012_05311655}, to molecular design~\cite{griffiths2020constrained}, to chip design~\cite{oh2022batch,oh2022bayesian} and quantum control~\cite{PRXQuantum.1.020322}.
In this work, we apply BO for the first time to quantum code design. Note that while BO often struggles in high dimensional spaces, we show here that by carefully choosing the kernel of the Gaussian process we can still scale our experiments to practically relevant sizes.
We make the following novel contributions:
\begin{figure*}[t]
\label{fig:workflow}
    \centering

\begin{tikzpicture}[>=Stealth]

  \node[draw, rounded corners, 
  inner sep=8pt] (A) at (90:3.5cm) {
    \begin{minipage}{4cm}
      \textbf{Fit surrogate}\\
      \begin{center}
      \includegraphics[scale=0.5]{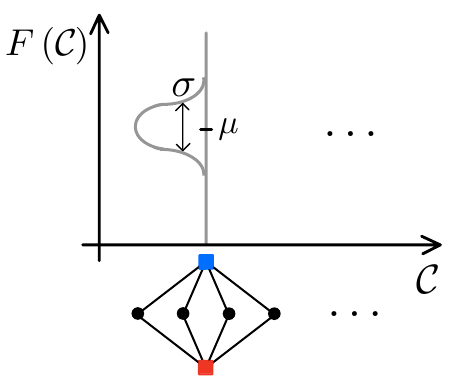}          
      \end{center}
    \end{minipage}
  };

  \node[draw, rounded corners, minimum width=4cm, minimum height=3cm, align=center] (B) at (330:3.5cm) {
    \begin{minipage}{4cm}
      \textbf{Optimize acquisition function}
      \begin{center}
      \includegraphics[scale=0.2]{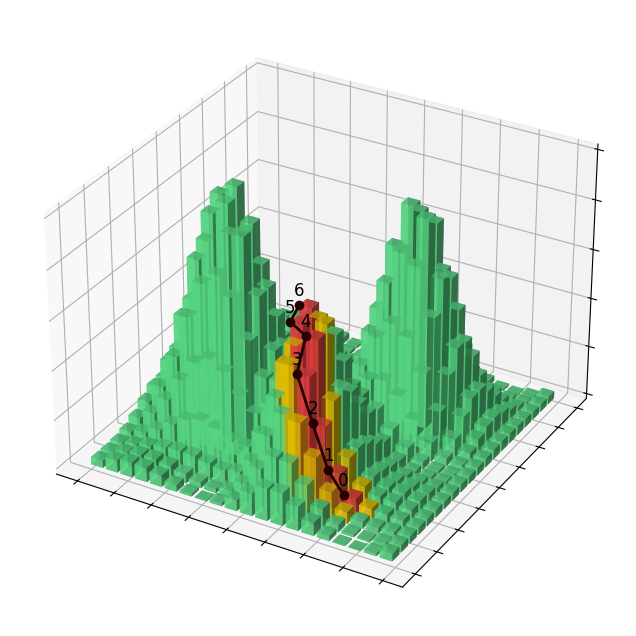}          
      \end{center}
  \end{minipage}
  };

  \node[draw, rounded corners, minimum width=4cm, minimum height=3cm, align=center] (C) at (210:3.5cm) {
    \begin{minipage}{4cm}
      \textbf{Evaluate code}\\
      \vspace{.1cm}
      \begin{center}
      \includegraphics[scale=0.45]{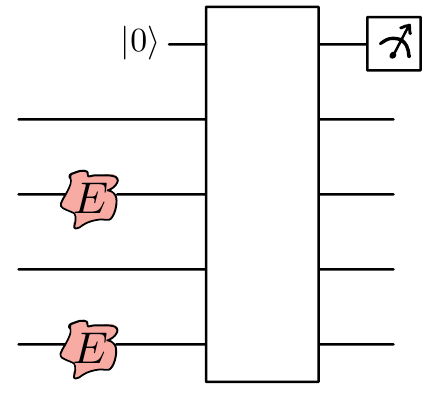}       
      \end{center}
      \vspace{.1cm}
  \end{minipage}
  };

  \draw[->, thick, bend left=12] (A) to (B);
  \draw[->, thick, bend left=12] (B) to (C);
  \draw[->, thick, bend left=12] (C) to (A);

\end{tikzpicture}
    \caption{The workflow of the Bayesian optimization algorithm.
    The algorithm iterates the following three steps until the computational budget is exhausted: (top) fit a probabilistic surrogate model to the obtained data to predict the objective function $F(\mathcal{C})$, e.g.~the logical error rate -- $\mathcal{C}$ denotes a quantum code represented as a Tanner graph and $\mu$ and $\sigma$ are the surrogate mean and uncertainty; (bottom right) optimize the acquisition function that uses the surrogate prediction and uncertainty 
    to propose a new candidate code -- here the figure depicts a hill climbing trajectory; (bottom left) evaluate the objective for the new code by performing a simulation of the code in the presence of errors $E$, which produces new data to fit the model to.
    }
    
\end{figure*}
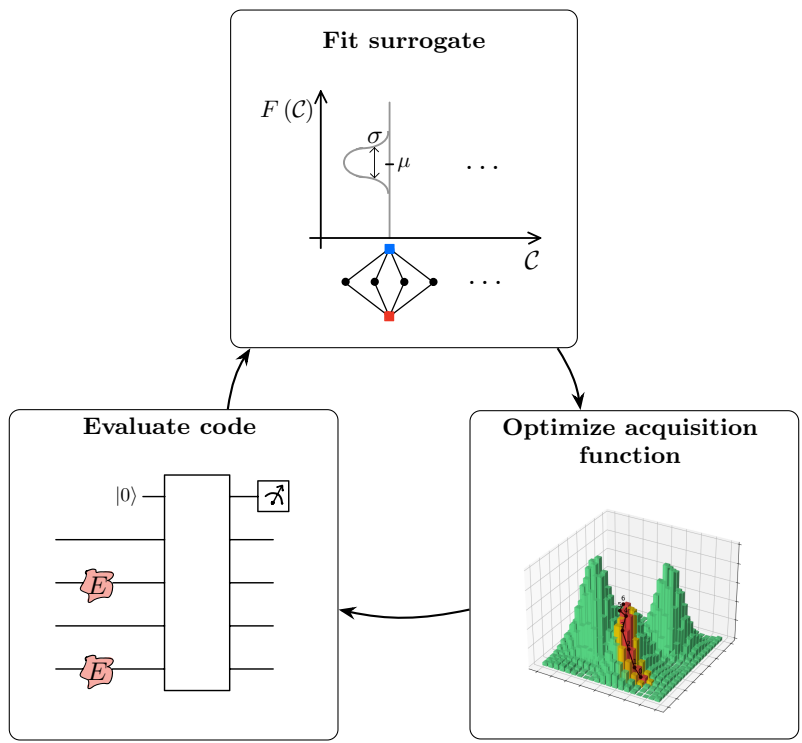
\begin{itemize}
    \item We introduce the \emph{first machine learning model that can accurately predict the logical error} of quantum LDPC codes under code capacity noise. The model is a graph neural network that uses message passing over multiple views of the chain complex representation of CSS codes. 
    Note that this is a different and a more challenging task than 
    decoding, for which 
     graph neural networks have been used previously, e.g. Ref.~\cite{maan2024machinelearningmessagepassingscalable}.
    \item We use this machine learning model as a probabilistic surrogate in a BO algorithm. 
    The workflow is summarized in Figure \ref{fig:workflow}.
    After evaluating initial points (i.e.~quantum codes), we fit the surrogate to the data and optimize 
    an acquisition function that balances exploration vs exploitation to pick the next point to evaluate. This loop continues for a predefined number of iterations, and at the end the best found code is returned.
    Our \emph{BO algorithm demonstrates faster convergence and superior final performance compared to evolutionary algorithms and random search}.
    \item 
    We report representative \emph{codes in the bivariate bicycle family discovered by BO}, including a high-rate 
    $[[144,36]]$ code and a low logical error rate per qubit
    $[[144,16]]$ code, both \emph{competitive with or surpassing established benchmarks} when optimized and tested under code capacity depolarizing noise.
\end{itemize}

Note that despite using a model of the objective function, BO retains all the benefits of model-free approaches in the literature, such as being able to adapt the solution to 
complex hardware constraints, including realistic noise models and connectivity constraints.
This is because the model in BO is learned and not handcrafted under simplistic assumptions.
Thus, our results pave the way to apply BO to such realistic hardware settings where the performance of a code is evaluated using an accurate simulator or directly on hardware.

The code to reproduce our experiments is available on GitHub at \url{https://github.com/Chengyuyihua/BO_for_QECcodes}.
\section{Preliminaries}

In this section, we review the basics of quantum error-correcting codes including the chain complex formalism and the evaluation of logical errors, and we detail how Bayesian optimization is applied to the problem at hand.

\label{SEC:pre}

\subsection{CSS codes and chain complex formalism}
\label{sec:css}
Calderbank–Shor–Steane (CSS) codes form an important subclass of stabilizer codes. 
In a stabilizer code, valid codewords $\ket{\Psi}$ are those quantum states that remain invariant under the action of all stabilizer operators -- that is, they are $+1$ eigenstates of elements of the stabilizer group $\mathcal{S}$, an Abelian subgroup of the $n$-qubit Pauli group $\mathcal{P}_n$ excluding the element $-I$.

In the CSS construction, the stabilizer group is generated by two commuting subsets of Pauli operators: the $X$-type and $Z$-type stabilizers,
denoted by $S_X^{(i)}$ and $S_Z^{(i)}$.
These are specified by two binary parity-check matrices
\[
H_x \in \mathbb{F}_2^{m_x \times n}, 
~~ 
H_z \in \mathbb{F}_2^{m_z \times n}, 
~~\text{where}~~ 
H_x H_z^{\top} = 0 .
\]
Each row of $H_x$ ($H_z$) defines an $X$-type ($Z$-type) stabilizer generator and the condition $H_x H_z^{\mathsf T} = 0$ ensures that $X$- and $Z$-stabilizers commute.
Explicitly, the $i$-th row of $H_x$ (or $H_z$) corresponds to
\[
S_X^{(i)} = \prod_{j=1}^{n} X_j^{\,H_x^{(i,j)}},
\qquad
S_Z^{(i)} = \prod_{j=1}^{n} Z_j^{\,H_z^{(i,j)}},
\]
where $H_x^{(i,j)}$ ($H_z^{(i,j)}$) denotes the $(i,j)$-th entry of $H_x$ ($H_z$), 
taking the value $1$ if the Pauli operator acts nontrivially on the $j$-th qubit and $0$ otherwise. 
The stabilizer group $\mathcal{S}$ is generated by all such $S_X^{(i)}$ and $S_Z^{(j)}$, i.e.
\[
\mathcal{S} = \langle S_X^{(1)},\dots,S_X^{(m_x)}, S_Z^{(1)},\dots,S_Z^{(m_z)} \rangle .
\]

Logical operators need to preserve the code space and are elements of the centralizer of the stabilizer group:
\[
C(\mathcal{S}) = \{ P \in \mathcal{P}_n \mid PS = SP \ \text{for all}\ S \in \mathcal{S} \},
\]
which commute with every element in $\mathcal{S}$.

The resulting $[[n,k,d]]$ CSS code, with length $n$, rate $k$, and distance $d$, encodes
\[
k = n - \mathrm{rank}(H_x) - \mathrm{rank}(H_z)
\]
logical qubits and corrects up to $t = \lfloor (d-1)/2 \rfloor$ errors, 
where the distance $d$ is the minimum weight of a nontrivial logical operator in $N(\mathcal{S}) / \mathcal{S}$.
As already remarked, computing the distance is hard in general
\cite{Kapshikar_2023}.

The algebraic structure of a CSS code can be expressed as a chain complex, namely as a sequence of vector spaces $C_i$ over $\mathbb{F}_2$
\begin{equation}
    \label{eq:chain}
C_2 \xrightarrow{\ \partial_2\ } C_1 \xrightarrow{\ \partial_1\ } C_0 ,
\end{equation}
together with boundary maps $\partial_i$, that are linear operators satisfying $\partial_1 \circ \partial_2 = 0$. 
The elements of $C_i$ are called $i$-chains and elements of the basis are called $i$-cells.
In our setting $C_1 = \mathbb{F}_2^n$ indexes physical qubits, while $C_2$ and $C_0$ index $Z$- and $X$-type stabilizers, respectively.
The boundary maps 
\begin{equation}
\label{eq:boundary}
    \partial_2 = H_z^{\top}: C_2 \to C_1, 
\quad 
\partial_1 = H_x: C_1 \to C_0 ,
\end{equation}
satisfy the chain-complex condition due to the commutativity conditions of $X$- and $Z$-stabilizers.

Logical operators correspond to nontrivial homology classes of this complex. 
$Z$-type logicals are represented by the first homology group
\begin{equation}
\label{eq:homology1}
    H_1(\mathcal{C}) = \ker(H_x)\big/ \mathrm{row}(H_z^{\top}),
\end{equation}
while $X$-type logicals are represented by the dual (cohomology) group
\begin{equation}
\label{eq:homology2}
    H_1^\vee(\mathcal{C}) = \ker(H_z)\big/ \mathrm{row}(H_x^{\top}) .
\end{equation}
Here, $\mathrm{row}(A)$ is the image of the linear map $A$.


As an example, consider the toric code. 
In this case, qubits are associated with edges (1-chains) of a two-dimensional cell complex. 
The $X$-type stabilizers correspond to vertices (0-chains), and the $Z$-type stabilizers correspond to faces or plaquettes (2-chains). 
The parity-check matrices 
$H_x, H_z$ correspond to the 
edge-vertex and edge-face incidence relations, respectively.
The $X$ and $Z$ logical operators correspond to non-trivial loops that wrap around the handles or holes of the surface, and the number of logical qubits is $2 = \dim H_1$. 
For more background on the chain complex representation of CSS codes, see Refs.~\cite{bombin2013introductiontopologicalquantumcodes,bravyi2014homological,cowtan2024css}.

The chain complex formalism underpins our embedding method described below, which maps the algebraic structure of $(H_x,H_z)$ into feature vectors using graph neural networks.
Another useful formulation of stabilizer codes is the Tanner graph representation, widely used for classical linear codes. The Tanner graph is a bi-partite graph which has vertices labeled by stabilizer generators and qubits, and has connections only between these two types of vertices, corresponding to non-zero entries of the parity check matrix.
Details of the code families used in the experiments below, bivariate bicycle and hypergraph product codes, can be found in appendices \ref{app:construction} and \ref{app:hgp}.


\subsection{Decoding and evaluation}

Decoding a CSS code amounts to inferring a likely error configuration from the measured syndrome. 
Formally, for parity-check matrices $H_x,H_z$, bit-flip and phase-flip errors $e_x,e_z\in\mathbb{F}_2^n$ yield syndromes
\[
s_x = H_z e_x, \qquad s_z = H_x e_z .
\]
A decoder aims to return an estimate $\hat e=(\hat e_x|\hat e_z)$ such that the combined error $e\oplus \hat e$ acts trivially on the code space. 
Finding the exact maximum-likelihood solution is \#P-complete for general stabilizer codes~\cite{sharpPcomplete}, motivating the use of heuristic or approximate algorithms.

Several decoders have been proposed for quantum LDPC codes, including belief propagation (BP) adapted from classical LDPC decoding, often enhanced with ordered-statistics postprocessing (BP-OSD)~\cite{Panteleev_2021,roffe2020decoding} or localized refinements (BP-LSD)~\cite{hillmann2024localized}. 
Such decoders achieve good performance across surface codes, hypergraph-product codes, and bicycle codes.

The performance of a decoder is evaluated by the logical error rate, defined as the probability that the combined actual and estimated error 
$e\oplus \hat e$ implements a nontrivial logical operator. 
Given a noise model, the logical error rate can be estimated via Monte Carlo sampling: simulate many random error instances, decode each, and record the fraction of failures. 
Reliable estimation typically requires $10^5$–$10^7$ trials for each candidate code, making evaluation of the logical error rate the dominant computational cost in 
an optimization algorithm that attempts to minimize it.

All logical error rates computed in this work employ the BP-LSD decoder~\cite{hillmann2024localized}, which scales well with block length.

\subsection{Bayesian optimization}
\label{sec:gpandbo}
We view the search for high-performing CSS codes as an expensive black-box optimization problem. 
Each candidate code $x$ is associated with an objective value $f(x)$, defined based on its logical error rate under a given decoder. 
Because direct evaluation requires costly Monte Carlo simulations, we employ Bayesian optimization (BO) with a Gaussian process surrogate to guide the search.
For definiteness, we assume that we want to maximize the objective $f$ in the following discussion.

From a probabilistic perspective, a Gaussian Process (GP) defines a prior distribution over functions \cite{rasmussen2003gaussian}.
A GP over a set $\mathcal{X}$ 
is defined by a mean function
$m:\mathcal{X}\to \mathbb{R}$
and 
a positive-definite 
kernel function $k:\mathcal{X}\times \mathcal{X}\to \mathcal{R}$.
The GP associates to inputs $X=\{x_1,\dots,x_N\}\in\mathcal{X}^N$ a multivariate Gaussian random variable with mean $m(X) = (m(x_1),\dots,m(x_N))$
and covariance 
$K(X,X)$ with matrix elements
$k(x_i,x_j)$.

Given data $\mathcal{D} = \{(x_i,y_i)\}_{i=1}^N$, GPs model 
the posterior predictive distribution at a test point $x$ as a Gaussian random variable
distributed according to  
the conditional probability of observing $f(x)$ given the data $\mathcal{D}$ \cite{rasmussen2003gaussian}:
\[
f(x)\,|\,\mathcal{D} \sim 
\mathcal{N}\big(\mu(x), \sigma^2(x)\big),
\]
where
\[
\begin{aligned}
\mu(x) &= m(x) + K(x,X)K(X,X)^{-1}\big(Y - m(X)\big), \\[4pt]
\sigma^2(x) &= k(x,x) - K(x,X)K(X,X)^{-1}K(X,x).
\end{aligned}
\]
Here $Y=(y_1,\dots,y_N)$ and 
$K(x,X) = (k(x,x_1),\dots,k(x,x_N))$.
The mean $\mu(x)$ serves as the Bayesian estimate of the function value, 
while the variance $\sigma^2(x)$ quantifies the model’s predictive uncertainty.
Popular kernels are the squared exponential
$k(x,x')=\sigma_f^2 \exp(-\|x-x'\|^2/2\ell)$ and the more general Matérn kernel, which depends on $r=\|x-x'\|$ as
\begin{equation}
\label{eq:matern}
k_{\nu}(r) = \sigma_f^2 \frac{2^{1-\nu}}{\Gamma(\nu)} \left( \frac{\sqrt{2\nu}r}{\ell} \right)^{\nu} K_{\nu}\left( \frac{\sqrt{2\nu}r}{\ell} \right)\,.
\end{equation}
Here $\nu > 0$ is the smoothness parameter, $\ell > 0$ is the length scale, $\sigma_f^2$ is the signal variance, $\Gamma$ is the gamma function, and $K_{\nu}$ is the modified Bessel function of the second kind.
See Appendix \ref{app:kernel} for more details on the Matérn kernel, which will be our choice in the experiments discussed below.

The GP surrogate informs an acquisition function $a(x)$, which selects new candidates by balancing exploitation (i.e.~finding point where $\mu(x)$ is small) and exploration (i.e.~points where $\sigma^2(x)$ is large).
We adopt the expected improvement (EI) criterion, defined with respect to the best observed value $f^\star$:
\[
    a(x) = \mathbb{E}\big[\max(0, f(x) - f^\star)\big],
\]
which has the closed-form expression
\[
a(x) 
=
\sigma(x)\left(z\Phi(z)+\phi(z)\right)
\,,
\]
where $\Phi(\cdot)$ and $\phi(\cdot)$ denote the standard normal cumulative distribution function and probability density function, respectively, and $z=(\mu(x)-f^*)/\sigma(x)$.

The BO procedure then iterates the following steps described also in Figure \ref{fig:workflow}:
\begin{enumerate*}
    \item Fit the GP surrogate to all available evaluations $\mathcal{D}$.  
    \item Approximately maximize $a(x)$ (e.g., by local search) to propose the next candidate $x_{\mathrm{new}}$.  
    \item Evaluate $f(x_{\mathrm{new}})$ via simulation to obtain the true logical error rate.  
    \item 
    Augment the dataset $\mathcal{D} \leftarrow \mathcal{D}\cup\{(x_{\mathrm{new}},f(x_{\mathrm{new}}))\}$ and return to step 1. 
\end{enumerate*}
This closed loop concentrates expensive evaluations on promising regions of the search space as captured by the surrogate, yielding a more sample-efficient discovery process than exhaustive or purely heuristic search.

Bayesian optimization has been applied in domains where the search space is combinatorial and evaluation is costly, such as 
molecular optimization 
\cite{korovina2020chembo},
permutation problems~\cite{oh2022batch} and VLSI macro placement~\cite{oh2022bayesian}. 
In such settings, acquisition maximization itself becomes challenging, and recent work has explored lightweight heuristics including hill climbing with random restarts~\cite{pmlr-v206-deshwal23a}. 
Inspired by these approaches, we adopt local search with potential restarts to optimize the acquisition function over the discrete space of CSS code configurations.

\section{Multi-view chain complex embedding for modeling logical error rate}
\label{SEC:MVCCE}


Accurate surrogate modeling of the logical error rate requires extracting informative structural features -- or embeddings -- of CSS codes. 
As remarked above, CSS codes admit a Tanner graph representation, which in fact 
underlies many decoders~\cite{caune2023belief,roffe2020decoding,hillmann2024localized}. 
This motivates both graph kernels, such as random walk kernels~\cite{kashima2003marginalized}, Weisfeiler-Lehman subtree kernels~\cite{shervashidze2009fast}, and graph neural networks~\cite{hamilton2017representation,jiang2019semi,xu2018powerful,yun2019graph}, which have demonstrated strong performance on graph data by capturing local and global graph features.  

However, the chain complex representation is a more appropriate representation of CSS codes than Tanner graphs, since it captures the commutativity relation between stabilizer generators. Further, work on lifted-product codes~\cite{panteleev2022asymptotically} establishes asymptotically good QLDPC codes by analyzing locally minimal (co)chains, highlighting the deep connection between code distance and local structure in chain complexes. 
These observations motivate the use of embeddings based on chain complexes that respect the algebraic structure of CSS codes.  

Recent advances in machine learning have extended graph representation learning to higher-order structures. 
Gaussian processes on cell complexes~\cite{alain2023gaussian} and neural networks on combinatorial complexes~\cite{hajij2022topological} provide frameworks that generalize graph kernels and message passing to richer topological spaces, capturing both local and global invariants. 
In parallel, topological architectures have been applied to molecular property prediction, where Molecular Hypergraph Neural Networks (MHNN)~\cite{chen2024molecular} leverage hypergraph structures to capture higher-order interactions, outperforming conventional GNNs. 

Motivated by the natural description of CSS codes as chain complexes and these advances in topological machine learning, we propose a \emph{multi-view chain complex embedding} tailored to CSS codes. 
Our approach first maps a code $\mathcal{C}$ to a multi-view representation derived from the chain complex induced by $(H_x,H_z)$. 
Using this representation we define a learnable embedding map
\[
\text{Emb}:\ \mathcal{Q}\to\mathbb{R}^{d_{f}}, \qquad z=\text{Emb}(\mathcal{C}),
\]
implemented by a multi-view message passing network described below. 
Here $\mathcal{Q}$ denotes the space of CSS codes.
We then place a GP prior on a latent function $g$ over the embedding space
to model the logical error rate $\widehat{p}_L$,
\[
g \sim \mathcal{GP}\!\big(m(\cdot),\, k(\cdot,\cdot)\big), \quad
\log \widehat{p}_L(\mathcal{C}) = g\big(\text{Emb}(\mathcal{C})\big),
\]
where $m$ and $k$ are the mean and kernel of the GP.
The posterior mean provides point predictions of the logical error rate, while the posterior variance supplies principled uncertainty for acquisition-based search. 

\subsection{Multi-view chain complex representation}
\label{sec:representation}

\begin{figure}[htbp]
    \centering
    \subfloat[Chain complex view]{%
        \includegraphics[width=.8\columnwidth]{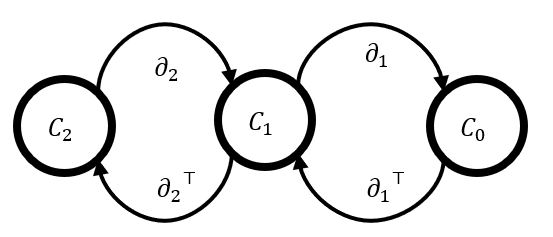}
    }\par
    \subfloat[Homology view]{%
        \includegraphics[width=.8\columnwidth]{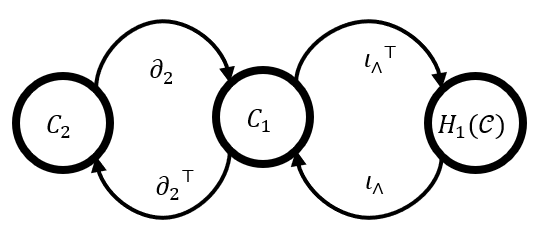}
    }\par
    \subfloat[Dual homology view]{%
        \includegraphics[width=.8\columnwidth]{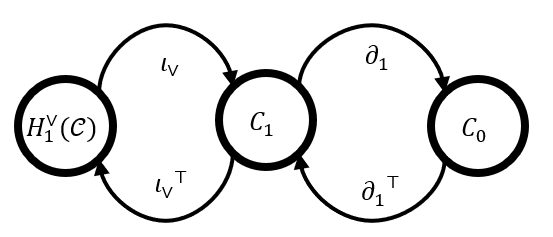}
    }
    \caption{Structure of the chain complex embedding. (a) Chain complex view: $C_i$ $(i=0,1,2)$ shown in circles are vector spaces over $\mathbb{F}_2$. Boundary mappings $\partial_i$ are represented by matrices over $\mathbb{F}_2$, and $\partial_i^\top$ denote the corresponding coboundary mappings. (b) Homology view: $H_1(\mathcal{C})$ is shown in a circle, with the mapping $\iota_\wedge: H_1(\mathcal{C}) \to C_1$ represented by a matrix. (c) Dual homology view: the dual counterpart of (b). In all views, circles represent vertex sets in a tripartite graph, and edges across parts correspond to their associated mappings. }
    \label{fig:multi_view}
\end{figure}

As introduced in Sec.~\ref{sec:css}, a CSS code $\mathcal{C}\in\mathcal{Q}$ admits a chain complex representation as in Eq.~\eqref{eq:chain}, with boundary maps specified in Eq.~\eqref{eq:boundary} and homology and cohomology groups given in Eq.~\eqref{eq:homology1} and \eqref{eq:homology2}.  

To make the action of logical operators explicit in the qubit space, we fix bases of $H_1(\mathcal{C})$ and $H_1^\vee(\mathcal{C})$ and select representative maps
\[
\iota_{\wedge} = L_z^\top: H_1(\mathcal{C}) \to C_1, 
\quad\!\!\!
\iota_{\vee} = L_x^\top: H_1^\vee(\mathcal{C}) \to C_1 ,
\]
where $L_z, L_x \in \mathbb{F}_2^{k\times n}$ are binary matrices whose rows are in $\ker(H_x)$ and $\ker(H_z)$, respectively. 
These maps inject logical classes into $C_1$, identifying concrete $1$-chains that act as logical operators on the physical qubits. 
The fact that $Z$ logical representatives yield no $X$ syndrome and vice versa,
is encoded in the following relations:
\[
H_x L_z^{\top} = 0, 
\qquad 
H_z L_x^{\top} = 0 \,.
\]

This structure motivates a \emph{multi-view representation} of a CSS code. 
An \emph{decode view} captures the tripartite incidence $\{C_2, C_1, C_0\}$ with edges given by $H_z$ and $H_x$, reflecting message passing as in standard syndrome decoding \cite{Panteleev_2021}. 
An \emph{$X$-logical view} augments $\{C_0, C_1\}$ with $H_1^\vee(\mathcal{C})$, connected via $\iota_{\vee}$, and a \emph{$Z$-logical view} augments $\{C_2, C_1\}$ with $H_1(\mathcal{C})$, connected via $\iota_{\wedge}$. 
These views incorporate not only one-hop Tanner graph adjacency but also two-step compositions and the quotient-level structure defined by homology. 
In this way the representation retains higher-order topology and algebra which, as we will show in the experiments, is essential for predicting the logical error rate from limited training data.

\subsection{Feature extraction}

\begin{figure}
    \centering
    \includegraphics[width=1\linewidth]{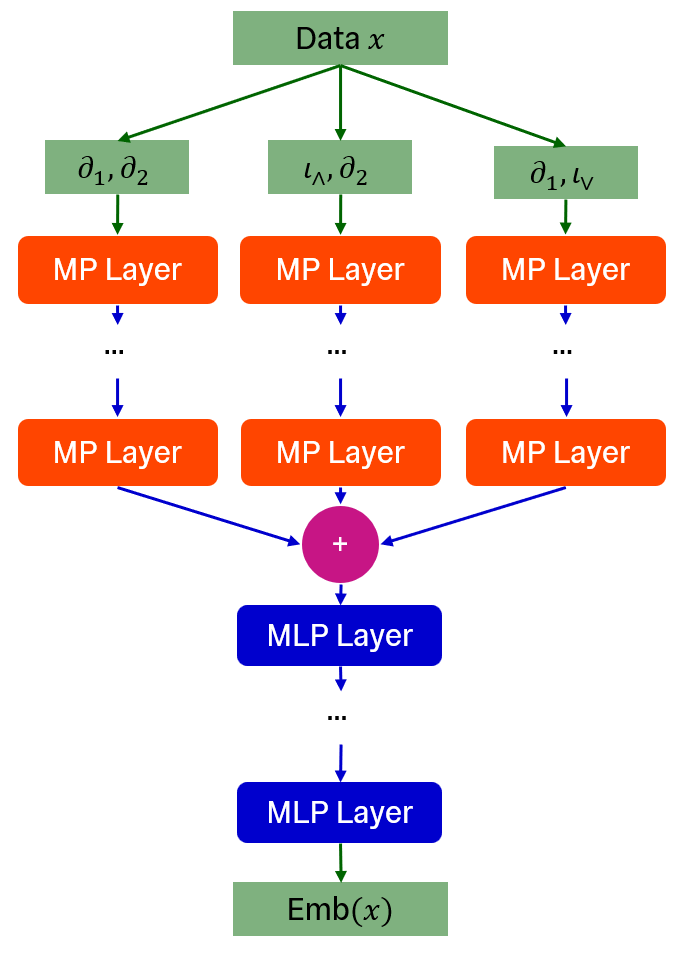}
    \caption{The architecture of the neural embedding $\operatorname{Emb}(x)$. MP Layer denotes a message-passing layer, and MLP Layer denotes a multi-layer perceptron. Given an input $x$, the data is first transformed into a multi-view chain-complex representation, consisting of the chain complex view ($\partial_1,\partial_2$), the homology view ($\iota_\wedge,\partial_2$) and the dual homology view ($\partial_1,\iota_\vee$). Each view is then processed independently by MP Layers to obtain a corresponding vector representation. These vectors are subsequently concatenated (depicted as $+$) and fed into a stack of MLP Layers. The output of the final MLP Layer is taken as the embedding Emb(x).}
    \label{fig:feat}
\end{figure}

The embedding procedure takes the boundary maps \(\partial_i\) and the inclusion maps \(\iota\) as input. The vector representations of the vertices in each part of the views are initialized with learnable type-specific embeddings, independent of relational structure. We then apply multiple message passing layers on each view, followed by multi layer perceptrons that operate on the concatenated view representations; this is summarized in Figure \ref{fig:feat}.

Let a relation \(\mathcal{N}:A\rightarrow B\) be given, where \(A,B \in \{C_0,C_1,C_2,H_1(\mathcal{C}),H_1^\vee(\mathcal{C})\}\)
are two of the spaces connected by edges in Figure \ref{fig:multi_view}
and 
\(\mathcal{N} \in \{\partial_1,\partial_1^\top,\ldots,\iota_\vee,\iota_\vee^\top\}\) depending on the choice of $A,B$.
This relation determines the direction of message passing.
For a vertex \(u \in B\), the neighborhood in $A$ is \(\mathcal{N}^\top(u)\). 

Representing each vertex \(u \in B\) by a vector \(x_u \in \mathbb{R}^{d_B}\), the message passing layer associated with relation \(\mathcal{N}\) is a mapping that given 
\[
\mathcal{MP}_{\mathcal{N}}:\ \mathbb{R}^{d_B} \rightarrow \mathbb{R}^{d'_B},
\]
that updates the value of the feature $x_u$ at $u$ via
\begin{equation}
\mathcal{MP}_{\mathcal{N}}(x_u)
= \phi_{\mathcal{N}}\!\left(
    x_u,\;
    \bigoplus_{v \in \mathcal{N}^\top(u)} f_{\mathcal{N}}(x_u, x_v)
\right),
\end{equation}
where \(\bigoplus\) denotes an aggregation operator, typically chosen as sum or mean. The functions \(f_{\mathcal{N}}(\cdot,\cdot)\) and \(\phi_{\mathcal{N}}(\cdot,\cdot)\) are learnable differentiable mappings -- which we take to be multilayer perceptrons -- that parameterize message computation and update, respectively. 
They are shared among all nodes for a given layer and their parameters vary across layers.

\begin{algorithm}[H]
\caption{Bayesian Optimization for Quantum Error-Correcting Code Discovery. Symbols are explained in the main text. 
}
\label{alg:BOonQEC}
 \hspace*{\algorithmicindent} \textbf{Input}:
 $\mathcal{X},
    F(\cdot),\text{Embedding}(\cdot|\theta_{\text{E}})$,\\
    \hspace*{1.75cm}$\mathcal{GP}(m(\cdot),k(\cdot,\cdot)\mid\theta_{\text{GP}})$, $a(\cdot)$, $T$
 \\
 \hspace*{\algorithmicindent} \textbf{Output}:
 $x_{\text{best}},y_{\text{best}}$
\begin{algorithmic}[1]
    \Procedure{BOonQEC}{}
        \State Initialize $X=\{x_1,\dots,x_{\nu_{0}}\},\;x_i\in\mathcal{X}$
        \State Evaluate $Y=\{y_1,\dots,y_{\nu_{0}}\},\;y_i=F(x_i)$
        \State $\mathcal{D}\gets\{(x_i,y_i)\}_{i=1}^{\nu_{0}}$; 
        \State fit GP parameters $\theta\gets\text{Fit}(\mathcal{GP},\text{Emb},\mathcal{D})$
        \For {$i \gets 1$ \textbf{to} $T$}
            \State $x_{\text{new}} \gets \text{OptimizerofAF}(\mathcal{GP},a(\cdot),\mathcal{X},\mathcal{D})$
            \State $y_{\text{new}} \gets F(x_{\text{new}})$
            \State $\mathcal{D}\gets \mathcal{D}\cup\{(x_{\text{new}},y_{\text{new}})\}$ and refit $\theta$
        \EndFor
        \State $i_{\text{best}}=\arg\max Y$
        \State \textbf{return} $x_{\text{best}}=X(i_{\text{best}})$, $y_{\text{best}}=Y(i_{\text{best}})$
    \EndProcedure

\end{algorithmic}
\end{algorithm}

\section{Bayesian Optimization for QEC codes}
\label{SEC:BO_for_QEC}

The workflow of our Bayesian optimization algorithm is shown in Figure~\ref{fig:workflow} and detailed in Algorithm~\ref{alg:BOonQEC}.
In Algorithm~\ref{alg:BOonQEC}, $\mathcal{X}$ denotes the discrete search space of code configurations. 
Each element $x\in\mathcal{X}$ represents a candidate quantum LDPC code.
During initialization, $\nu_0$ samples $\{x_i\}_{i=1}^{\nu_0}$ from $\mathcal{X}$ uniformly, and their corresponding performance values $y_i=F(x_i)$ are obtained by evaluating the black-box objective $F(\cdot)$. 
In Sec.~\ref{SEC:experiments} we will focus on codes parametrized by the lifted-product and hyper-graph product constructions and $F$ being a combination of code rate and logical error, but the algorithm is more general and applies to other code constructions and objective functions as well.

We define the surrogate model by combining the embedding function $\mathrm{Emb}(\cdot)$ 
and a Gaussian process $\mathcal{GP}(m(\cdot), k(\cdot,\cdot))$. 
As described in Secs.~\ref{SEC:MVCCE} and~\ref{sec:gpandbo}, 
the surrogate provides a predictive distribution for the objective value at any candidate $x$, 
conditioned on the observed dataset at time $t$
\[
\mathcal{D}_{t} = \{(x_i, y_i)\}_{i=1}^{\nu_0+t}.
\]
This predictive distribution forms the basis of the acquisition function used to select new evaluations.

Before running the optimization loop, and after each new data acquisition,
we tune the parameters of the embedding and the GP by maximizing the marginal likelihood of the observations:
\[
\theta_{\mathrm{E}}^{\ast}, \theta_{\mathrm{GP}}^{\ast}
= 
\argmax_{\theta_{\mathrm{E}},\,\theta_{\mathrm{GP}}} 
\; p\!\left(\{y_i\}_{i=1}^{v_t}\,\middle|\,\{x_i\}_{i=1}^{v_t}; 
\theta_{\mathrm{E}}, \theta_{\mathrm{GP}}\right),
\]
where $\theta_{\mathrm{E}}$ and $\theta_{\mathrm{GP}}$ denote the parameters of the embedding network 
and the Gaussian process (mean and kernel hyperparameters), respectively.  
This fitting procedure is standard in GP regression \cite{rasmussen2003gaussian} and performed here by gradient-based optimization methods.

At each iteration, the acquisition function $a(\cdot)$ (which is the expected improvement EI) proposes a new code candidate $x_{\text{new}}$ through the optimizer $\text{OptimizerofAF}$.

\begin{figure}[h]
    \centering
    \includegraphics[width=0.8\linewidth]{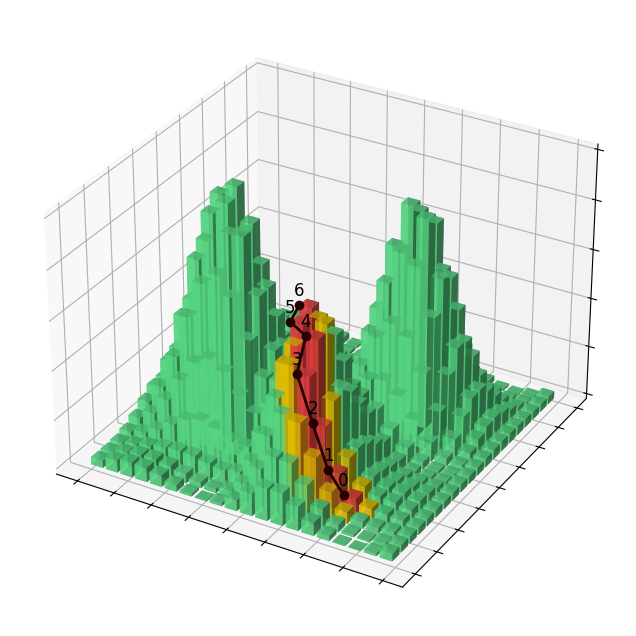}
    \caption{A sketch of hill climbing. The vertical axis represents the objective function. The black path indicates the sequence of selected points, with 0 denoting the starting point. Red bars highlight the best value in each iteration, while yellow bars mark the explored neighborhood. The algorithm terminates at step 6, which is returned as the final result.}
    \label{fig:hillclimbing}
\end{figure}

To optimize the acquisition function over the discrete search space 
$\mathcal{X}$ we employ a hill-climbing procedure. 
The algorithm requires a notion of neighborhood $\mathcal{N}(x)$ of a point $x\in\mathcal{X}$, which for the case of $\mathcal{X}$ being a space of binary vectors as considered in Sec.~\ref{SEC:experiments}, it is defined as the set of binary vectors differing from $x$ by exactly one bit.
Starting from a randomly selected point $x\in\mathcal{X}$, the hill climbing algorithm evaluates all configurations in $\mathcal{N}(x)$. The candidate with the largest acquisition value is chosen, and the process is repeated iteratively until no further improvement is found.

This local search strategy is particularly well suited for Bayesian optimization in discrete, high-dimensional spaces \cite{oh2022batch}. It efficiently exploits the structure of the acquisition landscape without requiring continuous relaxation, ensures that all updates remain feasible within the binary code representation, and leverages the typically smooth local behavior of the expected improvement function. Figure~\ref{fig:hillclimbing} illustrates a typical hill-climbing trajectory, where the procedure converges to a locally monotonic maximum of the acquisition function.

The iterative procedure of Bayesian Optimization continues for a total evaluation budget of $T$, yielding the best-performing configuration $x_{\text{best}}$ and its corresponding objective value $y_{\text{best}}$.
\section{Experiments and results}
\label{SEC:experiments}
\subsection{Code construction}

\label{sec:construct}
We consider bivariate bicycle (BB) codes constructed with the lifted-product (LP) formalism~\cite{panteleev2021quantum}
and hypergraph product codes \cite{tillich2013quantum}. Details on the constructions can be found in Appendices~\ref{app:construction} and \ref{app:hgp}.
In summary, we construct a BB code for the bivariate bicyclic group $G_\ell\times G_m$ (here $G_r$ is the cyclic group of order $r$) by specifying an element of the group algebra of this group using a binary representation $x \in \mathcal{X} = \mathbb{F}_2^{2(\ell+m-1)}$.
This allows us to represent each code as a point in a binary vector space, which facilitates the definition of the search space and the sampling of candidate codes in the subsequent experiments.
Similarly, a hypergraph product code is specified by two binary matrices $H_1\in\mathbb{F}_2^{m_1\times n_1}$ and $H_2\in \mathbb{F}_2^{m_2 \times n_2}$.
 
\subsection{Objective function}

To evaluate the performance of each sampled code, we estimate its logical error rate (LER) through Monte Carlo simulation under a fixed noise model and decoder. 
Specifically, we adopt the depolarizing noise model in the code capacity setting, where each qubit independently undergoes a Pauli error drawn uniformly from $\{X, Y, Z\}$ with a total probability of $p$. 
Decoding is performed using belief-propagation with the localized statistics decoding (BP-LSD) algorithm~\cite{hillmann2024localized}.

For each candidate code $\mathcal{C}$, $N$ random errors are sampled from the physical noise channel. 
The decoder is applied to each syndrome measured on the corrupted state, and the logical error rate is estimated as
\[
    p_{\mathrm{L}}(\mathcal{C}) = \frac{1}{N}\sum_{i=1}^{N} \delta_{\mathrm{L}}(e_i \oplus \hat{e}_i),
\]
where $\delta_{\mathrm{L}}(\cdot)$ equals $1$ if the residual error corresponds to a logical operation and $0$ otherwise.

For convenience of comparison across codes with different logical dimensions $k$, we also report the logical error rate per qubit (LERPQ), defined as the failure probability of a specific single qubit, under the assumption of independence of the failure probability over the logical qubits
\begin{equation}
\label{eq:wer}
    p_{\mathrm{PQ}} = 1 - (1 - p_{\mathrm{L}})^{1/k}.
\end{equation}

To relate empirical logical performance to theoretical limits, we introduce a pseudo-distance $\hat{t}$, 
which measures the average correctable error weight under the given noise model and decoder. 
It is obtained by inverting the binomial-tail approximation of the logical error rate:
\begin{align}
    p_{\mathrm{L}}
    &=
    \sum_{j=0}^{n} 
    g_j
    \binom{n}{j} p_p^{\,j} (1-p_p)^{\,n-j}
    \nonumber
    \\
\label{eq:psuedo_t}
    &\approx \sum_{j=\lfloor \hat{t} \rfloor+1}^{n} \binom{n}{j} p_p^{\,j} (1-p_p)^{\,n-j}.
\end{align}
Here 
$g_j$ is the fraction of weight-$j$ errors that lead to a failure and $p_p$ is the physical error rate on each qubit.
Intuitively, $\hat{t}$ represents the radius of a Hamming ball that would make the empirical logical performance 
consistent with the probability of exceeding $t$ physical errors under an i.i.d.~noise assumption. 
For inversion, we fit a low-degree polynomial $f_1(\hat{t})$ to $\log_2 p_L(\hat t)$, so that we can calculate a pseudo-distance $\hat{t} = f^{-1}_1(\log_2p_L)$.

For classical codes, the Hamming bound provides a fundamental tradeoff between distance and rate. 
For non-degenerate quantum stabilizer (CSS) codes correcting up to $t$ errors, the quantum Hamming bound~\cite{gottesman1997stabilizercodesquantumerror} takes the form
\begin{equation}
  2^{k}\sum_{j=0}^{t}\binom{n}{j}3^{j} \le 2^{n}.
\end{equation}
The Hamming bound for non-degenerate CSS codes is derived by counting Pauli error patterns of weight at most \(t\). 
If every logical codeword state \(c\) is surrounded by a Hamming ball
\[
B_t(c)=\{e:\operatorname{wt}(e)\le t\},
\]
and these balls are disjoint, then the total number of correctable patterns cannot exceed the size of the error space.  

Taking $\log_2$ and normalizing by $n$ yields
\begin{equation}
  \frac{k}{n} + \frac{1}{n}\log_2\!\Bigg(\sum_{j=0}^{t}\binom{n}{j}3^{j}\Bigg) \le 1.
\end{equation}
With the definition
\begin{equation}
\label{eq:psuedo_t}
  f_2(t) := \frac{1}{n}\log_2\!\Bigg(\sum_{j=0}^{t}\binom{n}{j}3^{j}\Bigg),
\end{equation}
the bound takes the simple additive form
\begin{equation}
  R + f_2(t) \le 1,
\end{equation}
where $R := k/n$ is the code rate.

We use the following objective function to quantify how closely a code approaches the quantum Hamming bound while accounting for its logical performance:
\begin{equation}
\label{eq:objective_function}
  F(x) = \lambda R_x + f_2\!\big(\hat t_x\big) - 1.
\end{equation}
By maximizing $F$ we produce codes that are close to saturating the Hamming bound. The weighting parameter $\lambda$ controls the tradeoff between proximity to the quantum Hamming bound and practical logical performance. In particular, $\lambda=1$ corresponds to ranking codes purely by their distance to the Hamming bound. By choosing $\lambda<1$, the objective function allows a moderate sacrifice in code rate in exchange for improved logical performance.
The pseudo-distance $\hat{t}$ plays the role of an effective average correctable weight: 
it represents the radius that makes this idealized ball-packing picture consistent with the empirical performance observed under the given channel and decoder.
While the quantum Hamming bound is derived strictly only for non-degenerate codes, it practically also holds for the codes we consider below \cite{10083270}.

\begin{table}[t]
\centering
\small
\setlength{\tabcolsep}{5pt}
\renewcommand{\arraystretch}{1.1}
\caption{The performance of different methods on the BB code test set. Here we compare the performance of GP on $(\ell,m)=(6,3)$ BB codes.
We evaluate predictive performance using mean squared error (MSE), coefficient of determination ($R^2$), and average negative log-likelihood (avg-NLL).
MSE measures pointwise accuracy; $R^2$ captures the fraction of variance explained by the model; and avg-NLL evaluates the quality of the full predictive distribution, combining accuracy and uncertainty calibration. Lower values of MSE and avg-NLL, and higher values of $R^2$, indicate better performance.}
\label{tab:main}
\begin{tabularx}{\linewidth}{>{\raggedright\arraybackslash}X p{0.13\linewidth} p{0.13\linewidth} p{0.13\linewidth}}

\toprule
\makecell[l]{\textbf{Method}} &
\textbf{MSE} &
\textbf{$R^2$} &
\textbf{avg-NLL} \\
\midrule
\makecell[l]{No embedding\\+ Matérn}             & 1.061 & -0.000 & 1.449 \\
\makecell[l]{No embedding\\+ WL kernel}          & 0.840 & 0.208  & 1.401 \\
\makecell[l]{GCN embedding\\+ Matérn}            & 0.822 & 0.226  & 1.358 \\
\makecell[l]{GIN embedding\\+ Matérn}            & 0.835 & 0.213  & 1.367 \\
\makecell[l]{GT embedding\\+ Matérn}             & 1.055 & 0.005  & 1.488 \\
\makecell[l]{1-view chain complex\\+ Matérn}     & 0.588 & 0.445  & 1.514 \\
\makecell[l]{3-view chain complex\\+ Matérn}     & 0.482 & 0.546 & 1.061 \\
\bottomrule
\end{tabularx}
\end{table}

\begin{figure}[htbp]
  \centering
  \subfloat[3-view chain complex embedding]{
    \includegraphics[width=1\columnwidth]{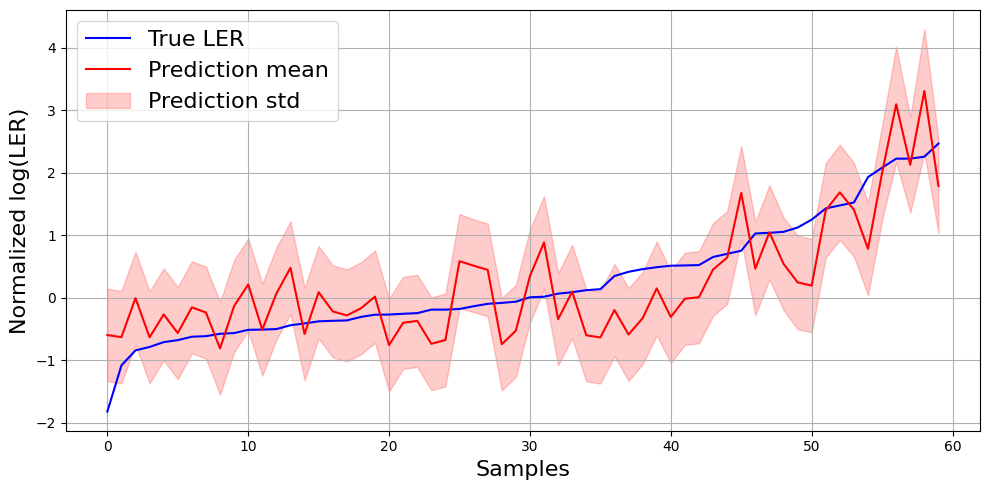}
  }\par\medskip
  \subfloat[1-view chain complex embedding]{
    \includegraphics[width=1\columnwidth]{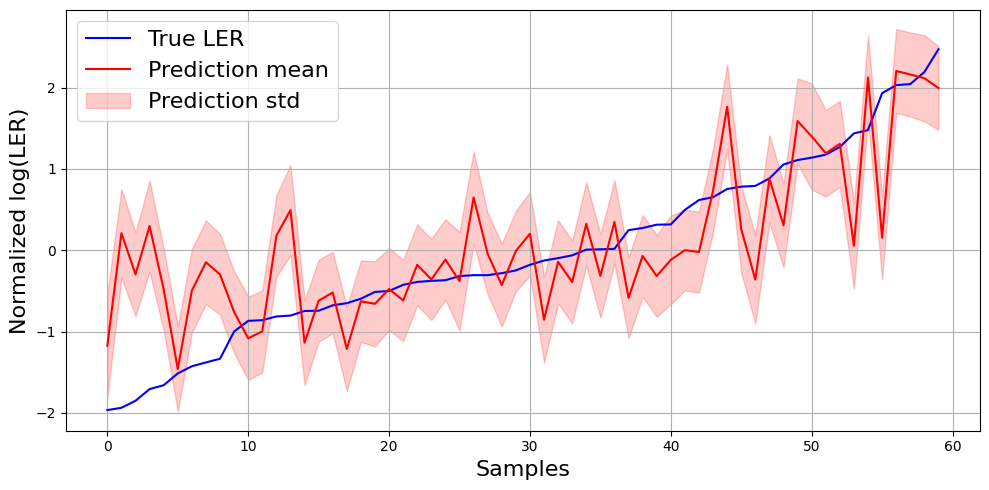}%
  }\par\medskip
  \subfloat[GCN embedding]{
    \includegraphics[width=1\columnwidth]{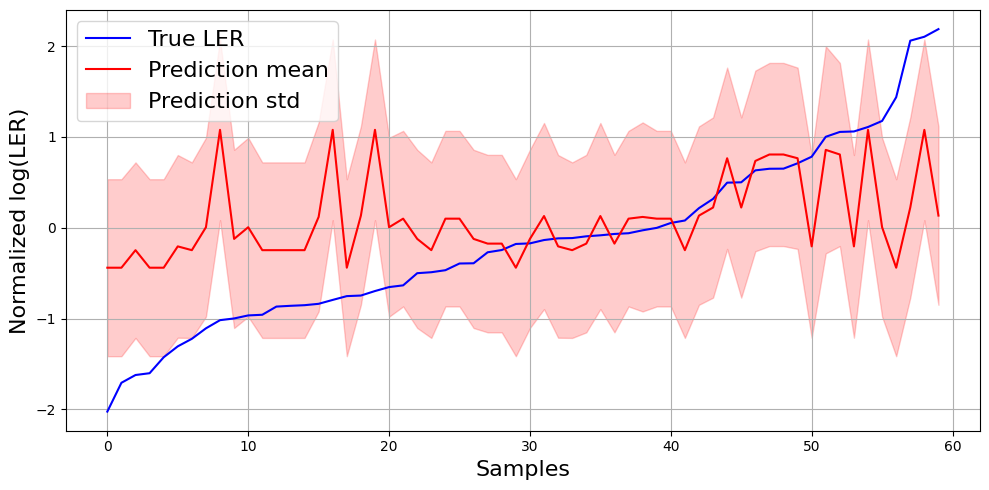}
  }
  \caption{The comparison between true values (blue line) and predicted mean $\pm\sigma$. The true LER values of all codes are shown by the blue curve and are sorted in ascending order from left to right.}
  \label{fig:comparison_embedding}
\end{figure}

\subsection{Embedding test}
\label{sec:embedding}

In this section, we evaluate how well the proposed multi-view chain complex embedding supports regression of the LER using a GP surrogate. The goal is to verify whether the embedding captures meaningful structural features of CSS codes that correlate with the logical error rate. We compare our approach against alternative embeddings and classical graph kernels.

All experiments use depolarizing noise ($p_p=0.05$)
in the code capacity setting decoded by BP-LSD; logical error rates are estimated via Monte Carlo sampling, and the standardized log-error serves as the GP regression target.
The GP employs a Matérn kernel~\eqref{eq:matern}, linear mean, and Gaussian likelihood, and is trained on the log-transformed target
\[
y = \log(p_L).
\]

To assess the effect of representation, we compare the proposed embedding against three baseline families under the same GP:
\begin{itemize}
    \item \textbf{No embedding}: a structure-agnostic baseline obtained by flattening the parity-check matrices \(H_x,H_z\).
    \item \textbf{WL kernel}: a classical Weisfeiler–Lehman subtree kernel applied to Tanner graphs, which uses 
    a hand-crafted structural similarity metric.
    \item \textbf{GNN embeddings}: learned Tanner-graph embeddings from GCN \cite{kipf2017semi}, GIN \cite{xu2019how}, and Graph Transformer 
    \cite{dwivedi2020generalization}
    encoders, coupled to the same GP.
\end{itemize}
Here, the WL kernel baseline is incorporated into the GP by replacing the Euclidean distance 
$\lVert x - x' \rVert$ in the Matérn kernel with the inner product between the WL subtree
feature vectors, $x^\top x',$ where $x,x'$ denote the fixed-dimensional WL subtree feature vector obtained from counting
relabeling patterns in the Tanner graph.

The proposed novel architectures tailored to quantum CSS codes are:
\begin{itemize}
    \item \textbf{1-view chain complex embedding}: Our representation where the entire CSS code -- namely the set of vertices corresponding to qubits, $X$-, $Z$-stabilizers, and logical operators -- is embedded as a single five-partite graph, with one message-passing layer on this graph.
    \item \textbf{3-view chain complex embedding}: Our full model, where qubits, stabilizers, and logical operators are separated into three distinct views (chain complex view, homology view, cohomology view), see Figure \ref{fig:feat}.
\end{itemize}
We remark that the difference between the 1-view chain complex embedding and the 3-view chain complex embedding is that in the former we do not separate the logical view and the chain complex view given by the parity check matrices. Instead, we put qubits, stabilizers, and logicals (or chains and (co)homology groups) in the same five-partite graph.

Monte Carlo cross-validation is performed using five repetitions of random 80/20 train-test splits, and the mean performance is reported in Table~\ref{tab:main}, where the dataset size is 300 BB codes randomly sampled and the training objective is the negative log marginal likelihood.

The no-embedding baseline provides a lower bound, as it ignores all structural information. The WL kernel incorporates handcrafted graph similarity but cannot adapt to the target task. Generic GNN embeddings improve upon both, confirming that local graph patterns correlate with logical reliability. The proposed chain complex embeddings outperform all alternatives: even the 1-view version performs strongly, indicating that it captures essential topological structure. The 3-view variant achieves the best across all metrics (MSE, $R^2$, and avg-NLL), showing improved accuracy and better-calibrated uncertainty.
Although the 1-view representation already encodes the full chain-complex structure, mixing qubits, stabilizers, and logical operators within a single message-passing graph forces the model to propagate all types of information through the same channels.
In contrast, the 3-view architecture respects the algebraic structure of CSS codes:
the chain view follows boundary relations,
the cochain view follows coboundary relations
and the logical view captures homology and cohomology classes.
Separating these dynamics prevents interference between structurally distinct relations and leads to representations that better preserve topological invariants.
Empirically, this yields improved predictive accuracy and reliability.

As a further validation of the proposed GP model of the logical error across quantum LPDC codes, we also investigated 
its performance for hypergraph product codes.
Specifically, we used 72-qubit BB codes with different \(l, m\) parameters and 72-qubit hypergraph product codes (see definition in Appendix~\ref{app:hgp}), each with 100 samples. As before, we used Monte Carlo cross validation and a 80/20 train/test split.
\begin{table}[t]
\hspace*{-1.8cm}  
\centering
\small
\setlength{\tabcolsep}{4pt}
\renewcommand{\arraystretch}{1.15}
\caption{Performance across different datasets. BB codes are sampled with $(\ell,m)\in \{(6,6),(12,3),(9,4)\}$. 
HGP codes are sampled with arbitrary $H\in \mathbb{F}_2^{6\times 6}$.}
\label{tab:main2}

\begin{tabularx}{0.96\linewidth}{   
    >{\raggedright\arraybackslash}p{1.6cm}
    >{\raggedright\arraybackslash}p{1.6cm} 
    *{3}{S[table-format=1.3]}
}
\toprule
\textbf{Embedding} & \textbf{Dataset} & \textbf{MSE} & \textbf{$R^2$} & \textbf{avg-NLL} \\
\midrule
\hline

\multirow{3}{*}{\makecell[l]{Proposed\\Embedding}}
  & 72-qubit BB codes  & 0.832 & 0.157 & 1.328 \\
  & 72-qubit HGP codes & 0.346 & 0.626 & 0.967 \\
  & BB + HGP codes     & 0.458 & 0.531 & 1.047 \\
\midrule
\hline

\multirow{3}{*}{\makecell[l]{GCN\\Embedding}}
  & 72-qubit BB codes  & 0.732 & 0.264 & 1.262 \\
  & 72-qubit HGP codes & 1.004 & -0.277 & 1.577 \\
  & BB + HGP codes     & 0.674 & 0.314 & 1.188 \\
\bottomrule
\end{tabularx}
\end{table}

The model shows slight variations in its ability to learn the error correcting ability of different code types, but overall, the embedding performs effectively across various datasets.

\subsection{The performance of Bayesian optimization}

\label{sec:bo}

In this section, we evaluate the performance of the proposed BO pipeline and compare its convergence behavior against evolutionary and random search baselines.

All experiments reuse the same CSS code family (BB codes based on the bivariate cyclic group of order $(\ell,m)$, see Appendix \ref{app:construction}), decoder, and depolarizing noise model described in the previous sections. 
Each candidate code $x$ is represented as a binary vector in the search space $\mathcal{X}=\mathbb{F}_2^{2(\ell+m-1)}$, 
and the simulator returns both the logical error rate $p_L(x)$ and the corresponding objective value $F(x)$ of Eq.~\eqref{eq:objective_function}.
A GP surrogate with a 3–view chain complex embedding is used as the probabilistic model. 

Neural network hyperparameters and priors follow the configuration detailed in Section~\ref{sec:embedding} and remain fixed across all optimization rounds.

The acquisition function is the expected improvement (EI). 

At each round, the Gaussian process provides predictive mean and variance of the LER for candidate codes. These predictions are propagated to obtain the corresponding mean and variance of the objective function via error propagation($\sigma_{g(x)}=|g'(x)|\sigma_x$). The resulting statistics are used to evaluate the EI, and new candidates are selected by the hill climbing algorithm.

We retain the previously optimized parameters and re-optimize the marginal likelihood after updating the training data. 

We compare BO to two reference optimizers under identical simulation conditions and evaluation budgets:
\begin{itemize}
    \item \textbf{Evolutionary algorithm (EA):} a population-based search that selects top individuals by objective value, applies bit-flip mutations and uniform crossover in Hamming neighborhoods, and carries elites across generations.
    \item \textbf{Random search (RS):} uniform sampling from $\mathcal{X}$ with identical validity constraints and evaluation budget.
\end{itemize}

Performance is reported as the best-so-far objective $\max_{i\le t} F(x_i)$ vs the evaluation index $t$.

\begin{figure}[htbp]
  \centering
  \subfloat[BB code, $n=36$ ($\ell{=}6$, $m{=}3$)]{
    \includegraphics[width=.8\columnwidth]{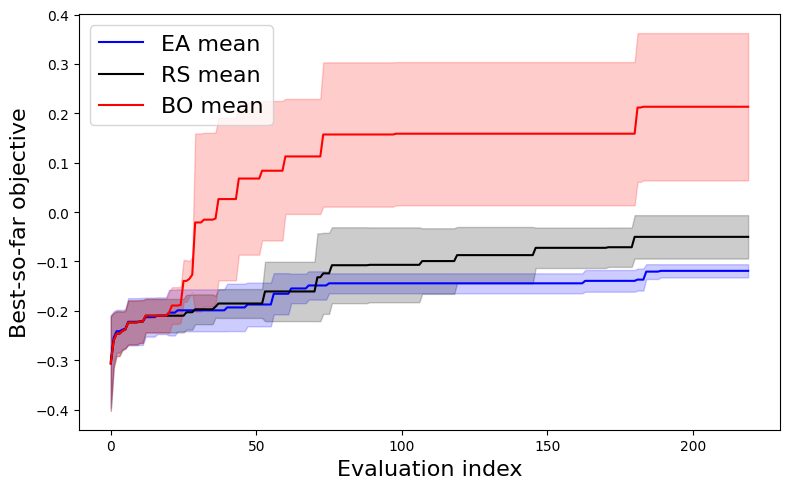}
  }\par\medskip
  \subfloat[BB code, $n=144$ ($\ell{=}12$, $m{=}6$)]{
    \includegraphics[width=.8\columnwidth]{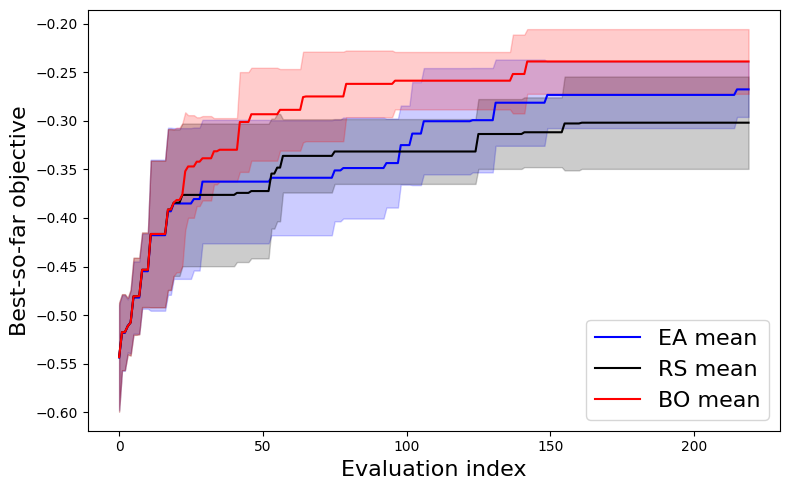}
  }
  \caption{Comparison of Bayesian Optimization and baseline methods. The parameters of the BB code search space are $(\ell,m)$ . The curves show the best-so-far values of the objective function evaluated during the optimization process. Solid lines indicate the mean performance across repeated runs, while the shaded regions represent $\pm \sigma$. In these experiments we set the parameter $\lambda=1$ in the objective function in Eq.~\ref{eq:objective_function}.}
  \label{fig:comparison_bo}
\end{figure}

Figure~\ref{fig:comparison_bo} shows that BO converges faster than both the evolutionary and random search baselines for $n=36$ and $n=144$. 
The best-so-far objective $\max_{i\le t} F(x_i)$ increases rapidly under BO, reaching higher final values within the same evaluation budget. 
This confirms that the surrogate model effectively guides exploration toward promising regions of the discrete search space.

The advantage arises from the combination of a structured embedding and a probabilistic surrogate. 
The chain complex embedding provides informative features of the code structure, while the GP posterior allows EI to focus sampling where improvement is most likely. 
The hill climbing inner loop translates local posterior gradients into feasible binary moves, increasing the chance of finding better codes in fewer evaluations. 
In contrast, evolutionary search relies on random mutations and selection without a global model, and random search lacks any exploitation mechanism.

Overall, these results demonstrate that the proposed BO framework achieves a higher sample efficiency and better final objectives than the tested evolutionary algorithm and purely random strategies. 
The framework applies broadly to any CSS code construction that can be encoded as a binary vector, and its scalability makes it suitable for large discrete design spaces where simulator evaluations are expensive.

\subsection{The performance of the discovered codes}
\label{sec:best}

\subsubsection{Logical error rate per qubit (LERPQ) performance}
\label{sec:LERPQperformance}
We compare discovered and reference codes using the LERPQ, which is defined in Eq.~\eqref{eq:wer}. All $p_L$ values are obtained with the same simulator, decoder, and noise model as specified earlier.

\begin{figure}
    \centering
    \includegraphics[width=\linewidth]{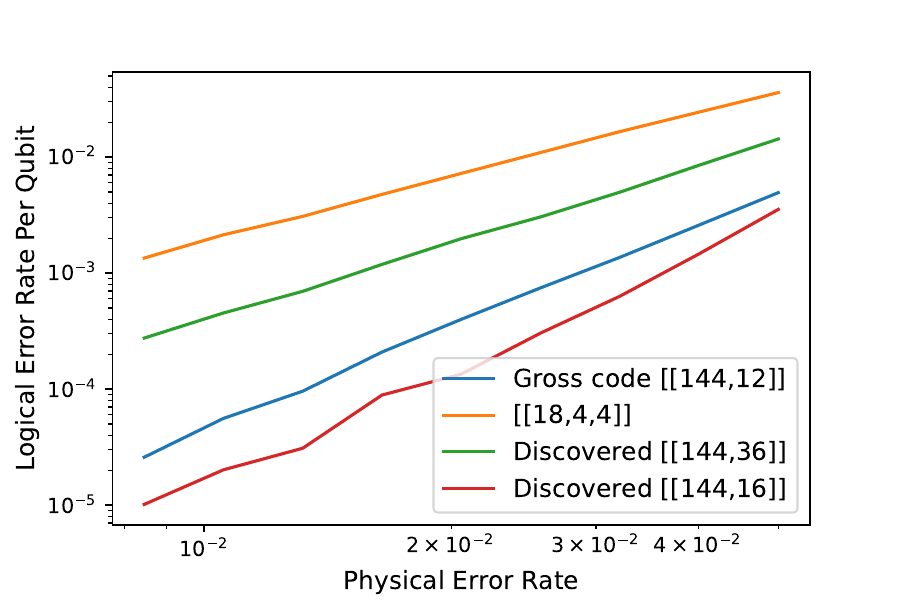}
    \caption{LERPQ versus physical error rate for the BO-discovered $[[144,36]]$ (with $\lambda=1$ objective function), a BO-discovered $[[144,16]]$ code (with $\lambda=0.5$) and for baselines $[[144,12,12]]$ and $[[18,4,4]]$. Lower is better. BO used the objective in Eq.~\eqref{eq:objective_function} with physical error rate $0.05$.}
    \label{fig:144code}
\end{figure}

Figure~\ref{fig:144code} reports LERPQ versus the physical error rate for two codes discovered by our proposed BO procedure at $(\ell,m)=(12,6)$: the $[[144,36]]$ code (with code rate 1/4) obtained by optimizing the $\lambda=1$ objective in Eq.~\eqref{eq:objective_function}, and the $[[144,16]]$ code obtained with $\lambda=0.5$. These are compared with the $[[144,12,12]]$ gross code baseline~\cite{bravyi2024high} and the $[[18,4,4]]$ bivariate bicycle code~\cite{wang2025demonstration} with a comparable rate (2/9).

The results reveal a clear rate-performance tradeoff.
Notably, the $[[144,16]]$ code simultaneously achieves a higher code rate and a consistently lower LERPQ than the $[[144,12,12]]$ gross code across the tested physical error range.

The $[[144,36]]$ code prioritizes code rate over error rate and achieves a three-fold increase relative to the gross code. This rate gain comes at the cost of a higher LERPQ; however, the degradation remains controlled and systematic over the explored regime. Moreover, at a code rate comparable to that of the $[[18,4,4]]$ code, the $[[144,36]]$ code shows a consistently lower LERPQ, indicating improved error correcting ability at similar rate.

\subsubsection{Code rate vs.\ $\hat{t}/n$ view}
\begin{figure}
    \centering
    \includegraphics[width=\linewidth]{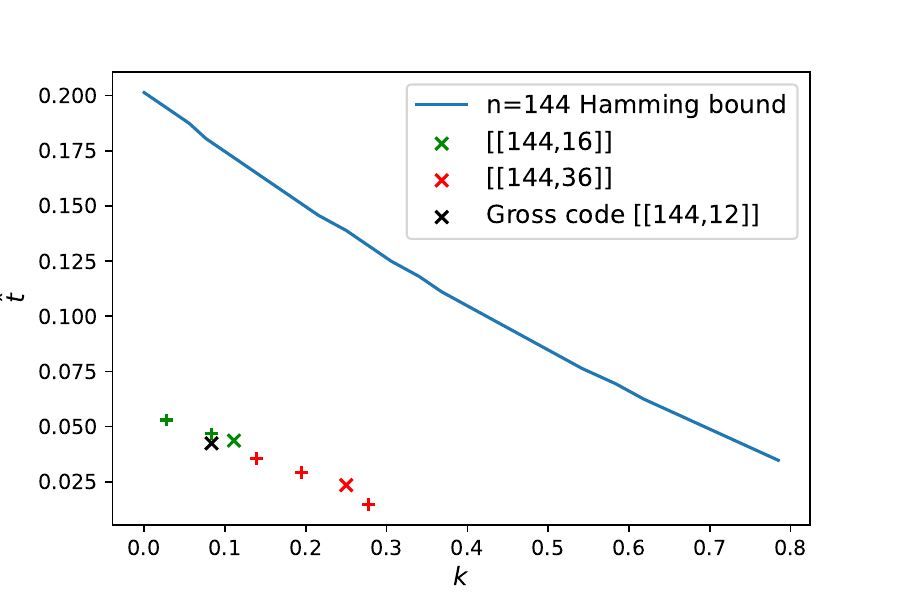}
    \caption{Operating points of BO-discovered bivariate bicycle codes at $n=144$ in the $(R,\hat t/n)$ plane, together with the non-degenerate CSS Hamming bound for $n=144$.
    The horizontal axis shows the code rate $R=k/n$, and the vertical axis shows the normalized pseudo-distance $\hat t/n$, where $\hat t$ is defined via the inversion in Eq.~\eqref{eq:psuedo_t}.
    Red markers denote codes discovered by optimizing the $\lambda=1$ objective in Eq.~\eqref{eq:objective_function}, which prioritizes proximity to the Hamming bound, while green markers denote codes discovered with $\lambda=0.5$. Cross markers highlight the representative codes discussed in the main text, including the $[[144,36]]$ and $[[144,16]]$ BO-discovered codes, as well as the $[[144,12,12]]$ gross code baseline.}
    \label{fig:hamming_bound}
\end{figure}

Figure~\ref{fig:hamming_bound} plots the non-degenerate CSS Hamming bound for $n=144$ together with the points of the BO-discovered bivariate bicycle codes in the $(R,\hat t/n)$ plane. This representation provides an objective-aligned view of the tradeoff between rate and effective correctable weight.

We emphasize that $\hat t$ is not an intrinsic invariant of a code: it depends on the decoder, the noise model, and the physical error rate at which it is evaluated. Moreover, code degeneracy and error correlations can invalidate a strict ball-packing interpretation. Nevertheless, $\hat t$ serves as a useful effective metric for comparing operating points under a fixed channel and decoding strategy.

All displayed points lie below the Hamming bound, as expected. Within $n=144$, the gross code $[[144,12,12]]$ serves as a baseline reference. Relative to this baseline, the BO-discovered codes are systematically shifted toward the Hamming bound in the $(R,\hat t/n)$ plane, indicating that the proposed optimization procedure succeeds in improving proximity to the theoretical rate-distance limit encoded in the objective.

We further observe that varying the objective weight $\lambda$ induces controlled biases along the rate-pseudo-distance tradeoff. Larger values of $\lambda$ favor operating points with higher code rate, whereas smaller values place greater emphasis on effective error-correcting capability as captured by $\hat t$. These tendencies are consistent with the LERPQ performance trends discussed in Sec.~\ref{sec:LERPQperformance}, for example in the behavior of the $[[144,36]]$ and $[[144,16]]$ codes, and illustrate how different optimization choices translate into distinct operating regimes relative to the Hamming bound.

\subsubsection{Discussion}
Across both plots, Bayesian optimization discovers bivariate bicycle codes under identical decoding and noise assumptions that improve upon strong baselines within a fixed cost function evaluation budget.

When the objective emphasizes proximity to the quantum Hamming bound, BO returns the $[[144,36]]$ code, which achieves a substantial increase in code rate relative to the $[[144,12,12]]$ gross code at the cost of reduced logical reliability as measured by LERPQ. When the objective prioritizes logical performance, BO returns the $[[144,16]]$ code, which outperforms $[[144,12,12]]$.
The corresponding rate vs $\hat{t}/n$ representation is consistent with these observations: in each case, the discovered codes move closer to the Hamming bound within the relevant normalized region, providing an objective-aligned consistency check for the observed LERPQ trends.

\section{Conclusion and Outlook}

This work introduces a Bayesian optimization framework for the discovery of CSS codes under fixed decoder and noise. 
This is based on a multi-view chain-complex neural embedding that enables a Gaussian process surrogate to predict logical performance from the code parity check matrix for the first time in the literature.
The Bayesian optimization algorithm achieves faster convergence and higher final objectives than evolutionary or random search baselines. 
When the objective balances accuracy and rate, Bayesian optimization discovers a $[[144,26]]$ code, which increases rate with respect to baselines while maintaining competitive logical error rate per qubit performance; when optimizing LERPQ alone it yields $[[144,8]]$, which surpasses baselines in the low-noise regime. 
These results demonstrate that surrogate-guided search can produce codes that approach the non-degenerate Hamming bound while remaining sample-efficient.

Our work paves the way towards applying our algorithm to real-world applications. This includes scaling to larger block lengths, searching broader code families such as multi-variate bicycle codes \cite{voss2025multivariatebicyclecodes}, incorporating circuit-level or correlated noise models, and exploring joint optimization over code and decoder parameters. 
In fact, our current surrogate performs comparatively poorly under circuit-level noise. This is because the embedding we use does not capture the temporal dynamics of the decoding graph, i.e. the time-extended Tanner graph used in circuit-level decoding.
Future work will address this by extending the embedding and surrogate to explicitly model spatio-temporal correlations in circuit-level noise.
Beyond these extensions, we envision the application of our framework to other important tasks towards fault-tolerance, such as designing quantum codes for distributed quantum computing or with optimized logical operations.
Finally, we anticipate that the multi-view chain complex embedding we discovered might be useful for designing neural decoders for quantum LDPC codes as well.

\begin{acknowledgments}
RB thanks 
E Egorov, 
A Pal, 
J Roffe and
M Welling for useful discussions and collaborations on related topics. RM acknowledges support from the EPSRC Grant number EP/W032643/1.
\end{acknowledgments}
\appendix

\section{Related work}
\label{app:related_work}

We here review related work that uses machine learning to optimize quantum codes.
Several ideas have been investigated:
Ref.~\cite{nautrup2019optimizing}
performs lattice modifications to a small toric code codes using reinforcement learning (RL) under code capacity noise, i.e.~when the noise affects qubits but not gates; 
Refs.~\cite{su2023discovery} and \cite{mauron2024optimization}
use RL to optimize distance and logical error rate of
tensor network codes with up to $17$ and $35$ qubits, respectively;
Ref.~\cite{olle2024simultaneous} optimizes encoding circuits of quantum codes 
with up to $\simeq 100$ qubits using RL
to directly impose the Knill-Laflamme conditions;  
Ref.~\cite{he2025discoveringhighlyefficientlowweight} minimizes stabilizer weights using RL, but the procedure requires distance computation at every step and is applicable only to moderate size hypergraph product codes;
Ref.~\cite{webster2024engineering} uses evolutionary algorithms to optimize the undetectable error rate of stabilizer codes up to $20$ qubits; 
Ref.~\cite{freire2025optimizing} 
fine tunes given hypergraph product codes using
simulated annealing and RL under erasure noise; 
Ref.~\cite{cao2022quantum}
uses variational and hybrid quantum-classical approaches to optimize codes with up to $14$ qubits;
Ref.~\cite{guerrero2025game} uses game-theory based strategies with a proxy distance as an objective without the possibility of adapting to different noise; 
Ref.~\cite{he2025co} uses 
multi-agent optimization powered by LLMs and demonstrations are up to $n=6$ qubits;
Ref.~\cite{lanka2025optimizingcontinuoustimequantumerror} directly searches for exponential subspaces in the continuous time error correction setting and is therefore limited to small systems.

\section{Code construction}
\subsection{Bivariate bicycle (BB) code construction}
\label{app:construction}

General bicycle (GB) codes~\cite{kovalev2013quantum} form an important subclass of QLDPC codes that can be naturally expressed in the lifted-product (LP) formalism.  
In the general GB construction, the $X$- and $Z$-check matrices are given by
\[
    H_x = [A \mid B], \qquad H_z = [B^{\mathsf{T}} \mid A^{\mathsf{T}}],
\]
where $A,B \in \mathbb{F}_2^{r\times r}$ satisfy
\[
    A B + B A = 0 .
\]
Typically, $A$ and $B$ are defined using structured or circulant matrices, which ensures that $H_x$ and $H_z$ inherit a high degree of regularity, enabling efficient encoding and decoding.

Following Bravyi \textit{et al.}~\cite{bravyi2024high}, BB codes can be viewed as a high-performance subclass of GB codes, obtained by setting
\[
 A = \mathbb{B}(a), \qquad
 B = \mathbb{B}(b),
\]
where $a,b \in \mathbb{F}_2 G$.  
Here, $\mathbb{F}_2 G$ denotes the group algebra of the bivariate cyclic group
\begin{align}
    G = G_\ell \times G_m, \quad 
    G_\ell = \langle g_\ell \rangle, \quad
    G_m = \langle g_m \rangle,
\end{align}
where $G_r = \langle g_r\rangle$ is the cyclic group of order $r$ generated by $g_r$,
so that
\begin{align}
    \mathbb{F}_2 G \cong \mathbb{F}_2[x,y]/(x^\ell-1,\, y^m-1).
\end{align}
Each generator $g_\ell, g_m$ corresponds to a circulant shift in the $\ell$ or $m$ dimension. Specifically, matrix representations $\mathbb{B}(g_\ell), \mathbb{B}(g_m)$ are defined as 
\begin{align}
\label{eq:cyclic}
\mathbb{B}(g_{\ell}) &= 
\left.
\begin{array}{c}
\overbrace{\begin{pmatrix}
0 & 1 & 0 & \cdots & 0 \\
0 & 0 & 1 & \cdots & 0 \\
\vdots & \vdots & \ddots & \ddots & \vdots \\
0 & 0 & \cdots & 0 & 1 \\
1 & 0 & \cdots & 0 & 0
\end{pmatrix}}^{\ell\ \text{columns}}
\end{array}
\right\}\!_{\ell\ \text{rows}}\!\otimes \mathbb{I}_m,
\\[6pt]
\mathbb{B}(g_{m}) &= 
\mathbb{I}_\ell \!\otimes\!
\left.
\begin{array}{c}
\overbrace{\begin{pmatrix}
0 & 1 & 0 & \cdots & 0 \\
0 & 0 & 1 & \cdots & 0 \\
\vdots & \vdots & \ddots & \ddots & \vdots \\
0 & 0 & \cdots & 0 & 1 \\
1 & 0 & \cdots & 0 & 0
\end{pmatrix}}^{m\ \text{columns}}
\end{array}
\right\}\!_{m\ \text{rows}}.
\end{align}

Notice that
\begin{align}
    n = 2\ell m, \quad k = \dim(\ker(A)\cap\ker(B))
\end{align}
For an arbitrary element $a \in G$, we define its binary vector representation as
\begin{align}
    \mathbbm{b}(a) = (\alpha_0, \dots, \alpha_{\ell+m-1}) \in \mathbb{F}_2^{\ell+m-1},
\end{align}
where
\begin{align}
    a = \alpha_0 e + \sum_{i=1}^{\ell-1} \alpha_i g_\ell^i + \sum_{i=\ell}^{\ell+m-1} \alpha_i g_m^{i-\ell} \in G,
    \quad \alpha_i \in \mathbb{F}_2.
\end{align}
The mapping $\mathbbm{b}(\cdot)$ thus provides a homomorphic correspondence between group elements and binary vectors.

When sampling candidate codes, we concatenate the binary representations of $a$ and $b$ as
\[
x = [\, \mathbbm{b}(a) \,\|\, \mathbbm{b}(b) \,] \in 
\mathcal{X} = \mathbb{F}_2^{2(\ell+m-1)}.
\]
This homomorphic embedding allows us to represent each BB code as a unique binary vector $x$, which facilitates the definition of a discrete search space for Bayesian optimization and efficient code sampling in the main experiments.
\subsection{Hypergraph product code construction}
\label{app:hgp}
Hypergraph Product (HGP) codes~\cite{tillich2013quantum} can be constructed from two arbitrary classical LDPC codes with parity-check matrices $H_1\in\mathbb{F}_2^{m_1\times n_1}$ and $H_2\in \mathbb{F}_2^{m_2 \times n_2}$.  The standard HGP form is
\begin{align}
    H_x = \big[ H_1 \otimes I_{n_2} \ \big|\ I_{m_1} \otimes H_2^{\mathsf{T}} \big], 
    \\[6pt]
    H_z = \big[ I_{n_1} \otimes H_2 \ \big|\ H_1^{\mathsf{T}} \otimes I_{m_2} \big].
\end{align}

This satisfies $H_x H_z^{\mathsf{T}} = 0$ and produces sparse parity-check matrices 
for the quantum code whose parameters are directly related to those of the classical LDPC codes:
\begin{align}
    n = n_1 n_2 + m_1 m_2,\quad k = k_1k_2 +  k_1^\top k_2^\top,
\end{align}
where $k_1 = \dim(\ker(H_1)),k_1^\top =  \dim(\ker(H_1^\top))$.
In our experiment in Sec.~\ref{sec:embedding}, we specify $H_1 = H_2\in \mathbb{F}_2^{6\times6}$.

\section{Kernels for the Gaussian process}
\label{app:kernel}

The Matérn class of covariance functions, see Eq.~\eqref{eq:matern}, is a widely used family of kernels that generalizes the squared exponential kernel by introducing a smoothness parameter $\nu$ \cite{rasmussen2003gaussian}. 

It controls the differentiability of the functions drawn from the Gaussian process. A sample function is mean-square differentiable $\lfloor \nu - 1 \rfloor$ times. In particular, for half integer values $\nu = p + \tfrac{1}{2}$ with integer $p \geq 0$, the kernel simplifies to an exponential times a polynomial. 
For $\nu \to \infty$ the Matérn kernel converges to the squared exponential kernel, corresponding to infinitely differentiable functions.

In practice the parameters of the Matérn kernel are adjusted to control different aspects of the function. The signal variance $\sigma_f^2$ sets the overall vertical scale of the function values, with larger values allowing the function to vary over a wider range. The length scale $\ell$ determines how quickly correlations decay with distance, so that small $\ell$ produces functions that vary rapidly while large $\ell$ produces smoother functions varying over longer ranges. As already remarked, the smoothness parameter $\nu$ controls the degree of differentiability, with small values such as $\nu=1/2$ yielding rough sample functions and larger values such as $\nu=5/2$ yielding smoother ones. 

The Matérn kernel is therefore more flexible than the squared exponential kernel because it can represent processes with different levels of smoothness.

\bibliographystyle{unsrtnat}
\bibliography{apssamp}

@article{panteleev2021quantum,
  title={Quantum LDPC codes with almost linear minimum distance},
  author={Panteleev, Pavel and Kalachev, Gleb},
  journal={IEEE Transactions on Information Theory},
  volume={68},
  number={1},
  pages={213--229},
  year={2021},
  publisher={IEEE}
}

@article{kovalev2013quantum,
  title={Quantum Kronecker sum-product low-density parity-check codes with finite rate},
  author={Kovalev, Alexey A and Pryadko, Leonid P},
  journal={Physical Review A—Atomic, Molecular, and Optical Physics},
  volume={88},
  number={1},
  pages={012311},
  year={2013},
  publisher={APS}
}

@article{shor1999polynomial,
  title={Polynomial-time algorithms for prime factorization and discrete logarithms on a quantum computer},
  author={Shor, Peter W},
  journal={SIAM review},
  volume={41},
  number={2},
  pages={303--332},
  year={1999},
  publisher={SIAM}
}

@article{tillich2013quantum,
  title={Quantum LDPC codes with positive rate and minimum distance proportional to the square root of the blocklength},
  author={Tillich, Jean-Pierre and Z{\'e}mor, Gilles},
  journal={IEEE Transactions on Information Theory},
  volume={60},
  number={2},
  pages={1193--1202},
  year={2013},
  publisher={IEEE}
}

@article{bravyi2024high,
  title={High-threshold and low-overhead fault-tolerant quantum memory},
  author={Bravyi, Sergey and Cross, Andrew W and Gambetta, Jay M and Maslov, Dmitri and Rall, Patrick and Yoder, Theodore J},
  journal={Nature},
  volume={627},
  number={8005},
  pages={778--782},
  year={2024},
  publisher={Nature Publishing Group UK London}
}

@article{webster2024engineering,
  title={Engineering Quantum Error Correction Codes Using Evolutionary Algorithms},
  author={Webster, Mark and Browne, Dan},
  journal={arXiv preprint arXiv:2409.13017},
  year={2024}
}

@article{nautrup2019optimizing,
  title={Optimizing quantum error correction codes with reinforcement learning},
  author={Nautrup, Hendrik Poulsen and Delfosse, Nicolas and Dunjko, Vedran and Briegel, Hans J and Friis, Nicolai},
  journal={Quantum},
  volume={3},
  pages={215},
  year={2019},
  publisher={Verein zur F{\"o}rderung des Open Access Publizierens in den Quantenwissenschaften}
}

@article{su2023discovery,
  title={Discovery of optimal quantum error correcting codes via reinforcement learning},
  author={Su, Vincent Paul and Cao, ChunJun and Hu, Hong-Ye and Yanay, Yariv and Tahan, Charles and Swingle, Brian},
  journal={arXiv preprint arXiv:2305.06378},
  year={2023}
}

@article{bausch2024learning,
  title={Learning high-accuracy error decoding for quantum processors},
  author={Bausch, Johannes and Senior, Andrew W and Heras, Francisco JH and Edlich, Thomas and Davies, Alex and Newman, Michael and Jones, Cody and Satzinger, Kevin and Niu, Murphy Yuezhen and Blackwell, Sam and others},
  journal={Nature},
  pages={1--7},
  year={2024},
  publisher={Nature Publishing Group UK London}
}

@misc{ruiz2024quantumcircuitoptimizationalphatensor,
      title={Quantum Circuit Optimization with AlphaTensor}, 
      author={Francisco J. R. Ruiz and Tuomas Laakkonen and Johannes Bausch and Matej Balog and Mohammadamin Barekatain and Francisco J. H. Heras and Alexander Novikov and Nathan Fitzpatrick and Bernardino Romera-Paredes and John van de Wetering and Alhussein Fawzi and Konstantinos Meichanetzidis and Pushmeet Kohli},
      year={2024},
      eprint={2402.14396},
      archivePrefix={arXiv},
      primaryClass={quant-ph},
      url={https://arxiv.org/abs/2402.14396}, 
}

@article{fosel2021quantum,
  title={Quantum circuit optimization with deep reinforcement learning},
  author={F{\"o}sel, Thomas and Niu, Murphy Yuezhen and Marquardt, Florian and Li, Li},
  journal={arXiv preprint arXiv:2103.07585},
  year={2021}
}

@misc{egorov2023end,
      title={The END: An Equivariant Neural Decoder for Quantum Error Correction}, 
      author={Evgenii Egorov and Roberto Bondesan and Max Welling},
      year={2023},
      eprint={2304.07362},
      archivePrefix={arXiv},
      primaryClass={quant-ph}
}

@article{Meinerz_2022,
   title={Scalable Neural Decoder for Topological Surface Codes},
   volume={128},
   ISSN={1079-7114},
   url={http://dx.doi.org/10.1103/PhysRevLett.128.080505},
   DOI={10.1103/physrevlett.128.080505},
   number={8},
   journal={Physical Review Letters},
   author={Meinerz, Kai and Park, Chae-Yeun and Trebst, Simon},
   year={2022},
   month=feb }

@article{Krastanov_2017,
   title={Deep Neural Network Probabilistic Decoder for Stabilizer Codes},
   volume={7},
   ISSN={2045-2322},
   url={http://dx.doi.org/10.1038/s41598-017-11266-1},
   DOI={10.1038/s41598-017-11266-1},
   number={1},
   journal={Scientific Reports},
   author={Krastanov, Stefan and Jiang, Liang},
   year={2017},
   month=sep }

@article{PhysRevLett.129.050507,
  title = {Optimal Control of Families of Quantum Gates},
  author = {Sauvage, Fr\'ed\'eric and Mintert, Florian},
  journal = {Phys. Rev. Lett.},
  volume = {129},
  issue = {5},
  pages = {050507},
  numpages = {7},
  year = {2022},
  month = {Jul},
  publisher = {American Physical Society},
  doi = {10.1103/PhysRevLett.129.050507},
  url = {https://link.aps.org/doi/10.1103/PhysRevLett.129.050507}
}

@article{niu2019universal,
  title={Universal quantum control through deep reinforcement learning},
  author={Niu, Murphy Yuezhen and Boixo, Sergio and Smelyanskiy, Vadim N and Neven, Hartmut},
  journal={npj Quantum Information},
  volume={5},
  number={1},
  pages={33},
  year={2019},
  publisher={Nature Publishing Group UK London}
}

@article{Gebhart_2023,
   title={Learning quantum systems},
   ISSN={2522-5820},
   url={http://dx.doi.org/10.1038/s42254-022-00552-1},
   DOI={10.1038/s42254-022-00552-1},
   journal={Nature Reviews Physics},
   publisher={Springer Science and Business Media LLC},
   author={Gebhart, Valentin and Santagati, Raffaele and Gentile, Antonio Andrea and Gauger, Erik M. and Craig, David and Ares, Natalia and Banchi, Leonardo and Marquardt, Florian and Pezzè, Luca and Bonato, Cristian},
   year={2023},
   month=feb }

@article{putterman2025hardware,
  title={Hardware-efficient quantum error correction via concatenated bosonic qubits},
  author={Putterman, Harald and Noh, Kyungjoo and Hann, Connor T and MacCabe, Gregory S and Aghaeimeibodi, Shahriar and Patel, Rishi N and Lee, Menyoung and Jones, William M and Moradinejad, Hesam and Rodriguez, Roberto and others},
  journal={Nature},
  volume={638},
  number={8052},
  pages={927--934},
  year={2025},
  publisher={Nature Publishing Group UK London}
}

@article{brock2025quantum,
  title={Quantum error correction of qudits beyond break-even},
  author={Brock, Benjamin L and Singh, Shraddha and Eickbusch, Alec and Sivak, Volodymyr V and Ding, Andy Z and Frunzio, Luigi and Girvin, Steven M and Devoret, Michel H},
  journal={Nature},
  volume={641},
  number={8063},
  pages={612--618},
  year={2025},
  publisher={Nature Publishing Group UK London}
}

@article{reichardt2024demonstration,
  title={Demonstration of quantum computation and error correction with a tesseract code},
  author={Reichardt, Ben W and Aasen, David and Chao, Rui and Chernoguzov, Alex and van Dam, Wim and Gaebler, John P and Gresh, Dan and Lucchetti, Dominic and Mills, Michael and Moses, Steven A and others},
  journal={arXiv preprint arXiv:2409.04628},
  year={2024}
}

@article{google2025quantum,
  title={Quantum error correction below the surface code threshold},
  journal={Nature},
  volume={638},
  number={8052},
  pages={920--926},
  year={2025},
  publisher={Nature Publishing Group UK London}
}

@article{mauron2024optimization,
  title={Optimization of tensor network codes with reinforcement learning},
  author={Mauron, Caroline and Farrelly, Terry and Stace, Thomas M},
  journal={New Journal of Physics},
  volume={26},
  number={2},
  pages={023024},
  year={2024},
  publisher={IOP Publishing}
}

@article{cao2022quantum,
  title={Quantum variational learning for quantum error-correcting codes},
  author={Cao, Chenfeng and Zhang, Chao and Wu, Zipeng and Grassl, Markus and Zeng, Bei},
  journal={Quantum},
  volume={6},
  pages={828},
  year={2022},
  publisher={Verein zur F{\"o}rderung des Open Access Publizierens in den Quantenwissenschaften}
}

@article{olle2024simultaneous,
  title={Simultaneous discovery of quantum error correction codes and encoders with a noise-aware reinforcement learning agent},
  author={Olle, Jan and Zen, Remmy and Puviani, Matteo and Marquardt, Florian},
  journal={npj Quantum Information},
  volume={10},
  number={1},
  pages={1--17},
  year={2024},
  publisher={Nature Publishing Group}
}

@article{roffe2020decoding,
  title={Decoding across the quantum low-density parity-check code landscape},
  author={Roffe, Joschka and White, David R and Burton, Simon and Campbell, Earl},
  journal={Physical Review Research},
  volume={2},
  number={4},
  pages={043423},
  year={2020},
  publisher={APS}
}

@article{caune2023belief,
  title={Belief propagation as a partial decoder},
  author={Caune, Laura and Reid, Brendan and Camps, Joan and Campbell, Earl},
  journal={arXiv preprint arXiv:2306.17142},
  year={2023}
}

@article{freire2025optimizing,
  title={Optimizing hypergraph product codes with random walks, simulated annealing and reinforcement learning},
  author={Freire, Bruno CA and Delfosse, Nicolas and Leverrier, Anthony},
  journal={arXiv preprint arXiv:2501.09622},
  year={2025}
}

@article{Gross_2006,
   title={Hudson’s theorem for finite-dimensional quantum systems},
   volume={47},
   ISSN={1089-7658},
   url={http://dx.doi.org/10.1063/1.2393152},
   DOI={10.1063/1.2393152},
   number={12},
   journal={Journal of Mathematical Physics},
   publisher={AIP Publishing},
   author={Gross, D.},
   year={2006},
   month=dec }

@misc{gottesman1997stabilizercodesquantumerror,
      title={Stabilizer Codes and Quantum Error Correction}, 
      author={Daniel Gottesman},
      year={1997},
      eprint={quant-ph/9705052},
      archivePrefix={arXiv},
      primaryClass={quant-ph},
      url={https://arxiv.org/abs/quant-ph/9705052}, 
}

@misc{lanka2025optimizingcontinuoustimequantumerror,
      title={Optimizing continuous-time quantum error correction for arbitrary noise}, 
      author={Anirudh Lanka and Shashank Hegde and Todd A. Brun},
      year={2025},
      eprint={2506.21707},
      archivePrefix={arXiv},
      primaryClass={quant-ph},
      url={https://arxiv.org/abs/2506.21707}, 
}

@article{dwivedi2020generalization,
  title={A Generalization of Transformer Networks to Graphs},
  author={Dwivedi, Vijay Prakash and Bresson, Xavier},
  journal={arXiv preprint arXiv:2012.09699},
  year={2020}
}

@inproceedings{xu2019how,
  title={How Powerful are Graph Neural Networks?},
  author={Xu, Keyulu and Hu, Weihua and Leskovec, Jure and Jegelka, Stefanie},
  booktitle={International Conference on Learning Representations (ICLR)},
  year={2019},
  note={arXiv:1810.00826}
}

@inproceedings{kipf2017semi,
  title={Semi-Supervised Classification with Graph Convolutional Networks},
  author={Kipf, Thomas N. and Welling, Max},
  booktitle={International Conference on Learning Representations (ICLR)},
  year={2017},
  note={arXiv:1609.02907}  
}

@ARTICLE{10083270,
  author={Dallas, Emanuel and Andreadakis, Faidon and Lidar, Daniel},
  journal={IEEE BITS the Information Theory Magazine}, 
  title={No $((n,K,d< 127))$ Code Can Violate the Quantum Hamming Bound}, 
  year={2022},
  volume={2},
  number={3},
  pages={33-38},
  doi={10.1109/MBITS.2023.3262219}}

@inproceedings{korovina2020chembo,
  title        = {{ChemBO}: Bayesian Optimization of Small Organic Molecules with Synthesizable Recommendations},
  author       = {Korovina, Ksenia and Xu, Shiliang and Kandasamy, Kirthevasan and Neiswanger, Willie and Poczos, Barnabas and Schneider, Jeff and Xing, Eric},
  booktitle    = {Proceedings of the 23rd International Conference on Artificial Intelligence and Statistics (AISTATS)},
  series       = {Proceedings of Machine Learning Research},
  volume       = {108},
  pages        = {3393--3403},
  year         = {2020},
  publisher    = {PMLR}
}

@article{Panteleev_2021,
   title={Degenerate Quantum LDPC Codes With Good Finite Length Performance},
   volume={5},
   ISSN={2521-327X},
   url={http://dx.doi.org/10.22331/q-2021-11-22-585},
   DOI={10.22331/q-2021-11-22-585},
   journal={Quantum},
   publisher={Verein zur Forderung des Open Access Publizierens in den Quantenwissenschaften},
   author={Panteleev, Pavel and Kalachev, Gleb},
   year={2021},
   month=nov, pages={585} }

@inproceedings{panteleev2022asymptotically,
  title={Asymptotically good quantum and locally testable classical LDPC codes},
  author={Panteleev, Pavel and Kalachev, Gleb},
  booktitle={Proceedings of the 54th annual ACM SIGACT symposium on theory of computing},
  pages={375--388},
  year={2022}
}

@article{PRXQuantum.1.020322,
  title = {Optimal Quantum Control with Poor Statistics},
  author = {Sauvage, Fr\'ed\'eric and Mintert, Florian},
  journal = {PRX Quantum},
  volume = {1},
  issue = {2},
  pages = {020322},
  numpages = {19},
  year = {2020},
  month = {Dec},
  publisher = {American Physical Society},
  doi = {10.1103/PRXQuantum.1.020322},
  url = {https://link.aps.org/doi/10.1103/PRXQuantum.1.020322}
}

@misc{maan2024machinelearningmessagepassingscalable,
      title={Machine Learning Message-Passing for the Scalable Decoding of QLDPC Codes}, 
      author={Arshpreet Singh Maan and Alexandru Paler},
      year={2024},
      eprint={2408.07038},
      archivePrefix={arXiv},
      primaryClass={quant-ph},
      url={https://arxiv.org/abs/2408.07038}, 
}

@misc{he2025discoveringhighlyefficientlowweight,
      title={Discovering highly efficient low-weight quantum error-correcting codes with reinforcement learning}, 
      author={Austin Yubo He and Zi-Wen Liu},
      year={2025},
      eprint={2502.14372},
      archivePrefix={arXiv},
      primaryClass={quant-ph},
      url={https://arxiv.org/abs/2502.14372}, 
}

@misc{voss2025multivariatebicyclecodes,
      title={Multivariate Bicycle Codes}, 
      author={Lukas Voss and Sim Jian Xian and Tobias Haug and Kishor Bharti},
      year={2025},
      eprint={2406.19151},
      archivePrefix={arXiv},
      primaryClass={quant-ph},
      url={https://arxiv.org/abs/2406.19151}, 
}

@misc{leverrier2022quantumtannercodes,
      title={Quantum Tanner codes}, 
      author={Anthony Leverrier and Gilles Zémor},
      year={2022},
      eprint={2202.13641},
      archivePrefix={arXiv},
      primaryClass={quant-ph},
      url={https://arxiv.org/abs/2202.13641}, 
}

@misc{bombin2013introductiontopologicalquantumcodes,
      title={An Introduction to Topological Quantum Codes}, 
      author={H. Bombin},
      year={2013},
      eprint={1311.0277},
      archivePrefix={arXiv},
      primaryClass={quant-ph},
      url={https://arxiv.org/abs/1311.0277}, 
}

@article{Kapshikar_2023,
   title={On the Hardness of the Minimum Distance Problem of Quantum Codes},
   volume={69},
   ISSN={1557-9654},
   url={http://dx.doi.org/10.1109/TIT.2023.3286870},
   DOI={10.1109/tit.2023.3286870},
   number={10},
   journal={IEEE Transactions on Information Theory},
   publisher={Institute of Electrical and Electronics Engineers (IEEE)},
   author={Kapshikar, Upendra and Kundu, Srijita},
   year={2023},
   month=oct, pages={6293–6302} }

@article{griffiths2020constrained,
  title={Constrained Bayesian optimization for automatic chemical design using variational autoencoders},
  author={Griffiths, Ryan-Rhys and Hern{\'a}ndez-Lobato, Jos{\'e} Miguel},
  journal={Chemical science},
  volume={11},
  number={2},
  pages={577--586},
  year={2020},
  publisher={Royal Society of Chemistry}
}

@inproceedings{NIPS2012_05311655,
 author = {Snoek, Jasper and Larochelle, Hugo and Adams, Ryan P},
 booktitle = {Advances in Neural Information Processing Systems},
 editor = {F. Pereira and C.J. Burges and L. Bottou and K.Q. Weinberger},
 pages = {},
 publisher = {Curran Associates, Inc.},
 title = {Practical Bayesian Optimization of Machine Learning Algorithms},
 url = {https://proceedings.neurips.cc/paper_files/paper/2012/file/05311655a15b75fab86956663e1819cd-Paper.pdf},
 volume = {25},
 year = {2012}
}

@article{oh2022bayesian,
  title={Bayesian optimization for macro placement},
  author={Oh, Changyong and Bondesan, Roberto and Kianfar, Dana and Ahmed, Rehan and Khurana, Rishubh and Agarwal, Payal and Lepert, Romain and Sriram, Mysore and Welling, Max},
  journal={arXiv preprint arXiv:2207.08398},
  year={2022}
}

@article{oh2022batch,
  title={Batch bayesian optimization on permutations using the acquisition weighted kernel},
  author={Oh, Changyong and Bondesan, Roberto and Gavves, Efstratios and Welling, Max},
  journal={Advances in Neural Information Processing Systems},
  volume={35},
  pages={6843--6858},
  year={2022}
}

@article{hillmann2024localized,
  title={Localized statistics decoding: A parallel decoding algorithm for quantum low-density parity-check codes},
  author={Hillmann, Timo and Berent, Lucas and Quintavalle, Armanda O and Eisert, Jens and Wille, Robert and Roffe, Joschka},
  journal={arXiv preprint arXiv:2406.18655},
  year={2024}
}

@article{shervashidze2009fast,
  title={Fast subtree kernels on graphs},
  author={Shervashidze, Nino and Borgwardt, Karsten},
  journal={Advances in neural information processing systems},
  volume={22},
  year={2009}
}

@inproceedings{kashima2003marginalized,
  title={Marginalized kernels between labeled graphs},
  author={Kashima, Hisashi and Tsuda, Koji and Inokuchi, Akihiro},
  booktitle={Proceedings of the 20th international conference on machine learning (ICML-03)},
  pages={321--328},
  year={2003}
}

@article{hamilton2017representation,
  title={Representation learning on graphs: Methods and applications},
  author={Hamilton, William L and Ying, Rex and Leskovec, Jure},
  journal={arXiv preprint arXiv:1709.05584},
  year={2017}
}

@inproceedings{jiang2019semi,
  title={Semi-supervised learning with graph learning-convolutional networks},
  author={Jiang, Bo and Zhang, Ziyan and Lin, Doudou and Tang, Jin and Luo, Bin},
  booktitle={Proceedings of the IEEE/CVF conference on computer vision and pattern recognition},
  pages={11313--11320},
  year={2019}
}

@article{xu2018powerful,
  title={How powerful are graph neural networks?},
  author={Xu, Keyulu and Hu, Weihua and Leskovec, Jure and Jegelka, Stefanie},
  journal={arXiv preprint arXiv:1810.00826},
  year={2018}
}

@article{yun2019graph,
  title={Graph transformer networks},
  author={Yun, Seongjun and Jeong, Minbyul and Kim, Raehyun and Kang, Jaewoo and Kim, Hyunwoo J},
  journal={Advances in neural information processing systems},
  volume={32},
  year={2019}
}

@article{alain2023gaussian,
  title={Gaussian processes on cellular complexes},
  author={Alain, Mathieu and Takao, So and Paige, Brooks and Deisenroth, Marc Peter},
  journal={arXiv preprint arXiv:2311.01198},
  year={2023}
}

@article{hajij2022topological,
  title={Topological deep learning: Going beyond graph data},
  author={Hajij, Mustafa and Zamzmi, Ghada and Papamarkou, Theodore and Miolane, Nina and Guzm{\'a}n-S{\'a}enz, Aldo and Ramamurthy, Karthikeyan Natesan and Birdal, Tolga and Dey, Tamal K and Mukherjee, Soham and Samaga, Shreyas N and others},
  journal={arXiv preprint arXiv:2206.00606},
  year={2022}
}

@article{chen2024molecular,
  title={Molecular hypergraph neural networks},
  author={Chen, Junwu and Schwaller, Philippe},
  journal={The Journal of Chemical Physics},
  volume={160},
  number={14},
  year={2024},
  publisher={AIP Publishing}
}

@InProceedings{pmlr-v206-deshwal23a,
  title = {Bayesian Optimization over High-Dimensional Combinatorial Spaces via Dictionary-based Embeddings},
  author = {Deshwal, Aryan and Ament, Sebastian and Balandat, Maximilian and Bakshy, Eytan and Doppa, Janardhan Rao and Eriksson, David},
  booktitle = {Proceedings of The 26th International Conference on Artificial Intelligence and Statistics},
  pages = {7021--7039},
  year = {2023},
  volume = {206},
  series = {Proceedings of Machine Learning Research},
  month = {25--27 Apr},
  publisher = {PMLR},
}

@inproceedings{bravyi2014homological,
  title={Homological product codes},
  author={Bravyi, Sergey and Hastings, Matthew B},
  booktitle={Proceedings of the forty-sixth annual ACM symposium on Theory of computing},
  pages={273--282},
  year={2014}
}

@article{cowtan2024css,
  title={CSS code surgery as a universal construction},
  author={Cowtan, Alexander and Burton, Simon},
  journal={Quantum},
  volume={8},
  pages={1344},
  year={2024},
  publisher={Verein zur F{\"o}rderung des Open Access Publizierens in den Quantenwissenschaften}
}

@article{guerrero2025game,
  title={Game-Theoretic Discovery of Quantum Error-Correcting Codes Through Nash Equilibria},
  author={Guerrero, Rub{\'e}n Dar{\'\i}o},
  journal={arXiv preprint arXiv:2510.15223},
  year={2025}
}

@article{he2025co,
  title={Co-Designing Quantum Codes with Transversal Diagonal Gates via Multi-Agent Systems},
  author={He, Xi and Lu, Sirui and Zeng, Bei},
  journal={arXiv preprint arXiv:2510.20728},
  year={2025}
}

@article{sharpPcomplete,
    title = {Hardness of decoding quantum stabilizer codes},
    author = {Iyer P, Poulin D},
    journal = {IEEE Transactions on Information Theory},
    year = {2015}

}

@incollection{rasmussen2003gaussian,
  title={Gaussian processes in machine learning},
  author={Rasmussen, Carl Edward},
  booktitle={Summer school on machine learning},
  pages={63--71},
  year={2003},
  publisher={Springer}
}

@article{wang2025demonstration,
  title={Demonstration of low-overhead quantum error correction codes},
  author={Wang, Ke and Lu, Zhide and Zhang, Chuanyu and Liu, Gongyu and Chen, Jiachen and Wang, Yanzhe and Wu, Yaozu and Xu, Shibo and Zhu, Xuhao and Jin, Feitong and others},
  journal={arXiv preprint arXiv:2505.09684},
  year={2025}
}

\end{document}